\documentclass[aps,twocolumn,prx,superscriptaddress,footinbib]{revtex4-1}
\usepackage{amsmath,amssymb,bm}
\usepackage{graphicx}
\usepackage{epstopdf}
\usepackage{latexsym}
\usepackage[normalem]{ulem} 
\usepackage[usenames,dvipsnames]{color}
\usepackage{hyperref}
\usepackage{natbib}
\begin{document}

\newcommand{\ns}{\mathcal{N}_{\mathrm{s}}}
\newcommand{\sinc}{\mathrm{sinc}}
\newcommand{\nb}{\mathcal{N}_{\mathrm{b}}}
\newcommand{\warn}[1]{{\color{red}\textbf{* #1 *}}}

\newcommand{\eqplan}[1]{{\color{blue}\textbf{equations:{ #1 }}}}

\newcommand{\figplan}[1]{{\color{blue}\textbf{figures: { #1 }}}}

\newcommand{\tableplan}[1]{{\color{blue}\textbf{tables: { #1 }}}}

\newcommand{\warntoedit}[1]{{\color{blue}\textbf{EDIT: #1 }}}

\newcommand{\warncite}[1]{{\color{green}\textbf{cite #1}}}

\newcommand{\Rev }[1]{{\color{black}{#1}\normalcolor}} 
\newcommand{\Com}[1]{{\color{red}{#1}\normalcolor}} 
\newcommand{\RobCom}[1]{{\color{Mahogany}{#1}\normalcolor}} 
\newcommand{\JohnCom}[1]{{\color{purple}{#1}\normalcolor}} 
\newcommand{\MGCom}[1]{{\color{red}{MG: #1}\normalcolor}} 

\newcommand{\mytitle}{Exploring adiabatic quantum dynamics of the Dicke model in a trapped ion quantum simulator}

%

\title{\mytitle}
\date{\today}

\author{ A. Safavi-Naini}
\thanks{These two authors contributed equally}
\affiliation{JILA, NIST and University of Colorado, 440 UCB, Boulder, CO 80309, USA}
\affiliation{Center for Theory of Quantum Matter, University of Colorado, Boulder, Colorado 80309, USA}
\author{ R. J. Lewis-Swan}
\thanks{These two authors contributed equally}
\affiliation{JILA, NIST and University of Colorado, 440 UCB, Boulder, CO 80309, USA}
\affiliation{Center for Theory of Quantum Matter, University of Colorado, Boulder, Colorado 80309, USA}
\author{J. G. Bohnet}
\affiliation{NIST, Boulder, Colorado 80305, USA}
\author{M. G\"{a}rttner}
\affiliation{JILA, NIST and University of Colorado, 440 UCB, Boulder, CO 80309, USA}
\affiliation{Center for Theory of Quantum Matter, University of Colorado, Boulder, Colorado 80309, USA}
\affiliation{Kirchhoff-Institut f\"{u}r Physik, Universit\"{a}t Heidelberg, Im Neuenheimer Feld 227, 69120 Heidelberg, Germany}
\author{K. A. Gilmore}
\affiliation{NIST, Boulder, Colorado 80305, USA}
\author{E. Jordan}
\affiliation{NIST, Boulder, Colorado 80305, USA}
\author{J. Cohn}
\affiliation{Department of Physics, Georgetown University, Washington, DC 20057, USA}
\author{J. K. Freericks}
\affiliation{Department of Physics, Georgetown University, Washington, DC 20057, USA}
\author{A. M. Rey}
\affiliation{JILA, NIST and University of Colorado, 440 UCB, Boulder, CO 80309, USA}
\affiliation{Center for Theory of Quantum Matter, University of Colorado, Boulder, Colorado 80309, USA}
\author{J. J. Bollinger}
\affiliation{NIST, Boulder, Colorado 80305, USA}

\begin{abstract}
We use a self-assembled two-dimensional Coulomb crystal of $\sim 70$ ions in the presence of an external transverse field to engineer a quantum simulator of the Dicke Hamiltonian. This Hamiltonian has spin and bosonic degrees of freedom which are encoded by two hyperfine states in each ion and the center of mass motional mode of the crystal, respectively. The Dicke model features a quantum critical point separating two distinct phases: the superradiant (ferromagnetic) and normal (paramagnetic) phases. We experimentally explore protocols that aim to adiabatically prepare the superradiant ground state, a spin-boson cat state with macroscopic phonon occupation, which is well-suited for enhanced metrology and quantum information processing. 
We start in the normal phase, with all spins aligned along a large transverse field and ramp down the field across the critical point following various protocols.  
We measure the spin observables, both experimentally and in our simulations to characterize the state of the system at the end of the ramp. We find that under current operating conditions an optimally designed ramp is not sufficient to achieve significant fidelity with the superradiant ground state. However, our theoretical investigation shows that slight modifications of experimental parameters, together with modest reductions in decoherence rates and thermal noise can increase the cat-state  fidelity to  $\sim 75 \%$ for  $N\sim20$ spins. Our results open a path for the use of large ensembles of trapped ions as powerful quantum sensors and quantum computers.
\end{abstract}

\maketitle  
\section{Introduction} 

Complex quantum manybody systems can be used to access entangled states that are a quantum resource for a broad range of applications. For example, cat-states are an excellent tool for quantum-enhanced metrology \cite{Leibfried2005,Gilchrist2004,Giovannetti1330,Strobel424}, and cluster states are an important resource for one-way quantum computation \cite{NIELSEN2006147}. Such entangled states can be generated through non-equilibrium dynamics or by adiabatic state preparation. Both approaches require the ability to dynamically control system parameters.

Quantum simulators with coupled spin and bosonic degrees of freedom are emerging as powerful quantum many-body platforms with easily tunable parameters.
These include cavity QED systems \cite{Leroux2010c,Ritsch2013} and trapped-ion arrays \cite{Porras2004,Kim2009}. In the latter case, the ion spins are coupled to one or multiple motional modes of the Coulomb crystal via a spin-dependent drive, resulting in phonon-mediated effective spin-spin interactions. 

Most often, these systems operate in a regime where the phonons \emph{only} facilitate the effective spin-spin coupling and do not enter the dynamics, allowing for their use as quantum simulators of spin models. Great progress has been realized in this effective spin-model regime, including a recent observation of dynamical phase transitions in a one-dimensional chain of $53$ ions~\cite{Zhang2017}. However, in the context of adiabatic state preparation, experiments have been limited to $< 20$ ions. Among these, one can highlight adiabatic preparation of the frustrated ground-state of the anti-ferromagnetic long-range Ising model \cite{Richerme2013,Richerme2013_PRA}, as well as a proof-of-principle preparation  of symmetry protected topological states in a spin-1 chain\cite{Senko2015}.

While there are notable exceptions for single trapped-ion experiments \cite{Pedernales2015,Lv2017,Johnson2017,Kienzler2016,Monroe1996,Toyoda2015,Debnath2017}, the regime where the full spin-boson dynamics has to be considered has remained largely unexplored in the context of many-ion quantum simulators.  In this work, we focus on this regime by realizing a trapped-ion quantum simulator of the Dicke model, an iconic model in cavity QED which describes the coupling of a (large) spin and an oscillator. The Dicke model is of broad interest as it exhibits rich physics including quantum phase transitions and non-ergodic behaviour. More recently it has gained renewed attention due to its implementation in cold-atom systems \cite{Baumann2010,Baumann2011,Klinder2015} and circuit QED \cite{Fink2009}.

Here, we investigate adiabatic quantum state transfer in an array of $\sim 70$ ions. The spin degree of freedom of the ions (encoded in the two lowest hyperfine states) are coupled via phonon-mediated long-range interactions induced by laser forces, tuned only to excite the center of mass (COM) mode of the Coulomb crystal. With the addition of an applied transverse field the experiment operates in a parameter regime where the phonons must be actively included for a proper description of the adiabatic protocol, which distinguishes our experiment from past efforts. We explore the performance of the simulator dynamics through various transverse field ramping schemes, and use theoretical calculations to benchmark the ground-state preparation fidelity and relevant spin observables, whilst also gaining insight into the relevant phonon dynamics. 
We also propose a simple procedure to disentangle the spin and phonon degrees of freedom of the targeted spin-boson cat-state, such that we can prepare a spin only cat-state.
Furthermore, we demonstrate that even in cases where the fidelity of the target state is not very high, one still can prepare  metrologically useful states.

\section{Spin-Boson System}

Our experimental system uses a single plane, two-dimensional (2D) array of laser-cooled
$^{9}$Be$^{+}$ ions in a Penning trap. The internal states forming the spin-1/2 system are the valence electron spin states in the ground state of $^{9}$Be$^{+}$ \cite{Martin2017_OTOC,Bohnet2016,Sawyer2012,Biercuk2009}. In the $4.46$~T magnetic field of the Penning trap, these states are split by $124$~GHz.
The interplay of the Coulomb repulsion  and  the  electromagnetic  confining  potentials creates a self-assembled ion crystal that  supports a set of normal modes \cite{Wang2013} which we couple to the spin degrees of freedom  via a spin-dependent optical dipole force (ODF), generated by the interference of a pair of detuned lasers with beatnote frequency $\omega_R$ \cite{Sawyer2012}.


We detune the ODF from the phonon modes such that the COM mode is predominantly excited and we uniformly couple all the ions in the crystal \cite{Bohnet2016}. Despite the fact that this regime limits the Hamiltonian to be collective, it allows us to generate highly entangled, and metrologically useful states while remaining within the bounds of computational feasibility for verification and benchmarking. In the presence of an additional transverse field, and in the frame rotating with $\omega_R$ the system is described by the Dicke Hamiltonian  \cite{Dicke1954,Garraway2011,Wall2017}:
\begin{eqnarray}
 \label{eq:HI} \hat{H}\left(t\right) & = & \hat{H}_{\mathrm{SB};I}+\hat{H}_B\left(t\right)+ \hat{H}_{\omega;I} ,\\
 \label{eq:HISB}\hat{H}_{\mathrm{SB};I} & = & -\frac{g_0}{\sqrt{N}} \left(\hat{a}+\hat{a}^{\dagger}\right)\hat{S}_z , \label{eqn:SpinPhCoupling} \\
 \label{eq:HField}\hat{H}_B & = & B(t) \hat{S}_x, \\
 \label{eq:HB} \hat H_{\omega;I} & = & - \delta \hat{a}^{\dagger}\hat{a}, 
\end{eqnarray}
where $\delta\equiv\omega_R-\omega_{\rm COM}$ and $\omega_{\rm COM}$ is the frequency of the COM mode. In this work $\delta$ is always negative. The operator $\hat{a} (\hat a^\dagger)$ is the bosonic annihilation (creation) operator for the COM mode, $B(t)$ is the time-varying strength of the applied transverse field, and $g_0$ represents the coupling between each ion and the COM mode. 
We have introduced the collective spin operators $\hat{S}_{\alpha} = (1/2)\sum_j \hat{\sigma}^{\alpha}_j$ where $\hat\sigma^{\alpha}_j$ is the corresponding Pauli matrices for $\alpha = x,y,z$ which acts on the $j$th ion.

\begin{figure*}[!]
 \includegraphics[width=\textwidth]{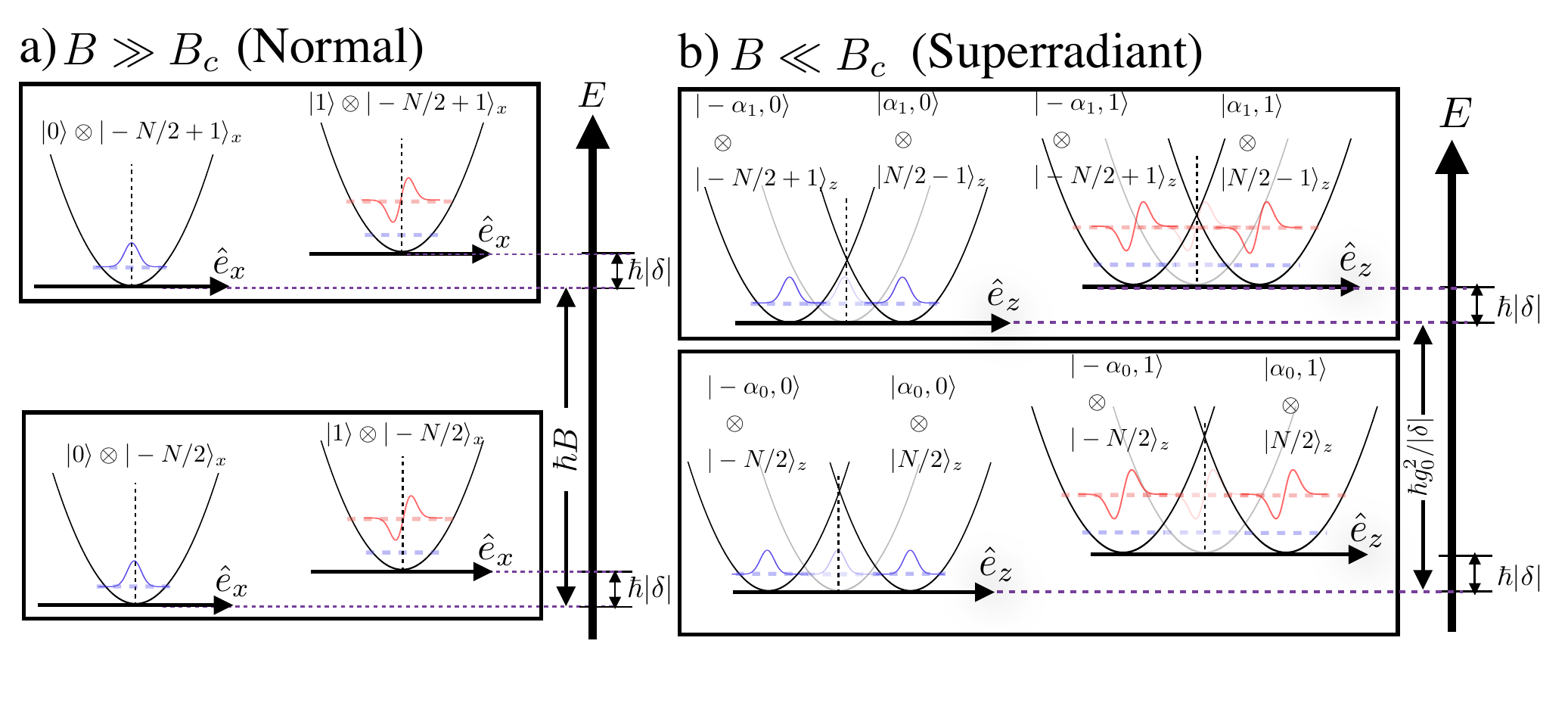}
 \caption{(a)-(b) Schematic of the low-lying energy levels of the Dicke Hamiltonian [Eq.~\eqref{eq:HI}] in two regimes, associated with the normal (paramagnetic) and superradiant (ferromagnetic) ground-state phases. (a) If the transverse-field is dominant then the ground-state is $\vert \psi_{\rm SB, B}^G\rangle$, the product of the vacuum phonon state, with all spins aligned along $B$. For $\delta<B$ the low lying excitations are phonon like. (b) If the spin-phonon coupling $\hat H_{\rm SB}$ is dominant the ground state is $\vert \psi_{{\rm SB},\, I}^G \rangle$. For $\delta^2<g_0^2$ the low lying excitations are displaced Fock states, here illustrated as the different occupied levels of two displaced harmonic oscillators. On the other hand, for $\delta^2 > g_0^2$ the excitations are spin-flips along $\hat{e}_z$. }
 \label{fig:schem}
\end{figure*}

The Dicke Hamiltonian exhibits a quantum phase-transition at $B_c=g_0^2/|\delta|$ in the thermodynamic limit \cite{Emary2003_PRL,Emary2003_PRE,Porras2013}, separating the normal and superradiant phases.  In the strong-field regime, $B \gg B_c$ (normal phase), the spins and phonons decouple into a product state. Since in this limit $\hat H_B$ is dominant, the lowest-energy configuration is for the spins to align against the field direction. We note that while we are always in the symmetric spin manifold ($S=N/2$), this state is often called a paramagnetic spin state, which is fully polarized with respect to the field with $\langle \hat{S}_x \rangle = -N/2$ and $\langle |\hat{S}_z| \rangle = 0$. The phonons occupy the vacuum state and the ground-state is given by,
\begin{equation}
 \vert \psi_{{\rm SB}, B} ^{GS}\rangle = \vert0\rangle \otimes \vert -N/2 \rangle_x , \label{eqn:SpinPhLargeB} 
\end{equation}
where we used  $\vert n \rangle$ for $n=0,1,...$ to denote Fock states and describe the spin degree of freedom in the basis $\vert S, M \rangle_x$ ($\vert S, M \rangle_z$) which are eigenstates of $\hat{S}^2$ and $\hat{S}_x$ ($\hat{S}_z$), labelled by their eigenvalue $S$ and $-S \leq M \leq S$ respectively. For brevity we contract the notation to $\vert S, M\rangle_x \equiv \vert M \rangle_x$ as we will only consider states in the fully symmetric manifold $S=N/2$. We also adopt the convention $\vert \psi_{SB} \rangle = \vert \psi_{\mathrm{phonon}} \rangle \otimes \vert\psi_{\mathrm{spin}} \rangle$ to describe the spin-phonon state, while the additional subscript $B$ is used to denote that this is the ground-state in the regime dominated by the transverse field $B$.

Conversely, in the weak-field limit $B \ll B_c$ (superradiant phase), the spin and phonon degrees of freedom are entangled. Here, the spin ground-state is ferromagnetic (or a linear combination of spins all up and all down) with $\langle |\hat{S}_z| \rangle = N/2$ and $\langle \hat{S}_x \rangle = 0$, and in the limit of $B \to 0$ is the entangled cat-state of spins \emph{and} phonons,
\begin{equation}
\vert \psi_{{\rm SB},\, I}^{GS} \rangle = \frac{1}{\sqrt{2}}\Big(\vert\alpha_0,0\rangle \otimes \vert N/2 \rangle_z \pm \vert-\alpha_0,0\rangle \otimes \vert -N/2 \rangle_z\Big),  \label{eqn:SpinPhCat}
\end{equation}
where the subscript $I$ denotes this ground-state is dominated by the spin-boson coupling.
Here, the phonon mode acquires a macroscopic occupation $\vert\alpha_0\vert^2$, where $\alpha_0 = g_0\sqrt{N}/(2\delta)$. The two states with spins all-up and all-down that make up the cat-state are separated by an energy gap at finite $B$ which is suppressed exponentially with atom number. We have introduced the displaced Fock states $\vert\alpha,n\rangle \equiv \hat{D}(\alpha)\vert n\rangle$ to characterize the phonons, with $\hat{D}(\alpha) = e^{\alpha\hat{a}^{\dagger} - \alpha^*\hat{a}}$ the associated displacement operator \cite{Wunsche1991}, which for $\alpha=0$ are equivalent to the previously defined Fock states $\vert 0, n\rangle \equiv \vert n \rangle$.

Finally, we note that the Dicke Hamiltonian has a spin-phonon parity symmetry such that $\hat{H}(t)$ is preserved under the simultaneous transformation of $\hat S_z\to -\hat S_z$, $\hat{S}_y \to -\hat{S}_y$ and $\hat a \to -\hat a$. The associated conserved quantity of the Hamiltonian is the generator of the symmetry $e^{i\pi(\hat{a}^{\dagger}\hat{a} + \hat{S}_x)}$. In the context of state preparation, this symmetry dictates that when ramping from high to low field, the state $\vert \psi_{{\rm SB},\, B}^{GS} \rangle$ will adiabatically connect to the superposition $\vert \psi_{{\rm SB},\, I}^{GS} \rangle$, to conserve the parity $\langle e^{i\pi(\hat{a}^{\dagger}\hat{a} + \hat{S}_x)} \rangle = e^{i\frac{\pi N}{2}}$. Specifically, for even $N$ the ground-state will be the symmetric superposition with $\langle e^{i\pi(\hat{a}^{\dagger}\hat{a} + \hat{S}_x)} \rangle = \pm 1$, whilst for odd $N$ the ground-state is the anti-symmetric superposition with $\langle e^{i\pi(\hat{a}^{\dagger}\hat{a} + \hat{S}_x)} \rangle = \pm i$. In contrast, this symmetry would be violated if one were to populate either of the states $\vert N/2\rangle_z \otimes \vert \alpha_0\rangle$ or $\vert -N/2\rangle_z \otimes \vert -\alpha_0\rangle$ (which have $\langle e^{i\pi(\hat{a}^{\dagger}\hat{a} + \hat{S}_x)} \rangle = 0$).

To construct efficient adiabatic protocols it is important to understand the low-lying excitations and energy spectrum of the Dicke Hamiltonian, and in particular how the interplay of phonon and spin degrees of freedom affects these. Both the strong- and weak-field ground-states, as well as their respective low-lying excitations  are shown schematically in Fig.~\ref{fig:schem}. The nature of the excitations are dictated by the relative magnitude of $\delta$, $g_0$, and $B$. 

In the large $B$ limit, if $\vert \delta \vert >B$ then the lowest energy excitations are spin-flips along $\hat e_x$. Conversely the excitations are phonon like if $\vert \delta \vert <B$.

In the weak-field limit, $B\ll B_c$, and if $\delta^2 < g_0^2$ the low-lying excitations are
phonon-like. In this case the lowest energy excited states correspond to displaced Fock states, $\vert \pm\alpha_0, n \rangle$ [see Fig.~\ref{fig:schem}(b)]. On the other hand, if $\delta^2>g_0^2$ then the low-lying excited states differ from the ground-state by spin-flips along $\hat e_z$.

The spin-component of the ground-states in the two regimes given by Eqs.~(\ref{eqn:SpinPhLargeB}) and (\ref{eqn:SpinPhCat}) strongly resemble the corresponding ground-states of the Lipkin model (LM) \cite{Morrison2008}, which emerges in the far-detuned regime wherein the phonons can be adibatically eliminated. 
This correspondence can be demonstrated by first rewriting the spin-boson Hamiltonian in the form: 
\begin{equation}
 \hat{H}(t) = -\delta \hat{b}^{\dagger}\hat{b} + \frac{J}{N} \hat{S}_z^2 + B(t)\hat{S}_x, \label{eqn:SpinBoson_H}
\end{equation}
where $\hat{b} = \hat{a} - [g_0/(\sqrt{N}\delta)]\hat{S}_z$ and $J = g_0^2/\delta$. The first term in the Hamiltonian describes phonons in a displaced potential, with a spin-dependent displacement $\alpha=[g_0/(\sqrt{N}\delta)]\hat{S}_z$, which accounts for the coherent occupation of the phonon mode in the weak-field ground-state [Eq.~(\ref{eqn:SpinPhCat})].

From Eq.~(\ref{eqn:SpinBoson_H}) we observe that in the limit of large detuning, $|\delta| \gg g_0/\sqrt{N},B$, the phonons can be adiabatically eliminated and we are left with an effective, spin-only Hamiltonian:
\begin{align}
\label{eq:HTFIM}
 \hat{H}_{\mathrm{LM}}(t) = \frac{J}{N} \hat{S}_z^2 + B(t)\hat{S}_x.
\end{align}
Here, we will always consider the case where $\delta < 0$ ($J<0$), which generates the ferromagnetic Lipkin model \cite{Elliott1970}. The Lipkin Hamiltonian exhibits a quantum critical point at $B_c=|J|\equiv g_0^2/|\delta|$, analogous to the Dicke Hamiltonian, which  separates the regimes of paramagnetically and ferromagnetically ordered spin ground-states.

While the ground-state physics of the Dicke Hamiltonian mirrors that of the LM in the strong and weak-field limits (in terms of the spin degree of freedom), their energy spectra can be markedly different depending on the choice of $\delta$. 

In the limit $|\delta| \gg g_0/\sqrt{N},B$ the eigen-spectrum of the two Hamiltonians is identical. However, as the magnitude of $\delta$ is decreased the phonon and spin degrees of freedom must be treated on an equal footing. Furthermore, at $B\simeq \vert \delta \vert$, the model features a resonance, where the states $\vert m \rangle\vert -N/2 \rangle_x $ and $\vert m-k\rangle\vert -N/2 + k \rangle_x$ , with $k$ a positive integer, become nearly degenerate and can be resonantly coupled. In fact the position of this resonance relative to the critical field strength $B_c$ allows us to categorize the effect of phonons on the overall dynamics into two cases: 
\begin{itemize}
\item {Case (i): $|\delta| \gg B_c$.} In this regime the resonance $B\simeq |\delta|$ lies well above the critical point. The ground-state $\vert \psi^{GS}_{SB; B} \rangle = \vert 0 \rangle\vert -N/2 \rangle_x$ is decoupled from other states at resonance. Thus, the dynamics can be affected by resonant couplings to other states (see above) only if excited states have become occupied throughout the evolution, for instance due to non-adiabaticity of the ramp.

\item{
Case (ii): $|\delta| \sim B_c$.} In this regime the resonance and the critical point are not well separated. As a result, the low-lying excitations at the critical point of the Dicke Hamiltonian are a non-trivial superposition of spin and phonon excitations, which is very different to those of the spin-only LM. A radical consequence of this complex interplay is the reduction of the energy gap between the ground and the first non-degenerate excited states at the critical point (referred to as the main gap throughout the paper), relative to the LM gap. This reduction is a crucial effect in the context of adiabatic state preparation, as the characteristic time-scale to remain adiabatic is inversely proportional to the energy gap. 
\end{itemize}

\begin{figure}[!]
 \includegraphics[width=0.5\textwidth]{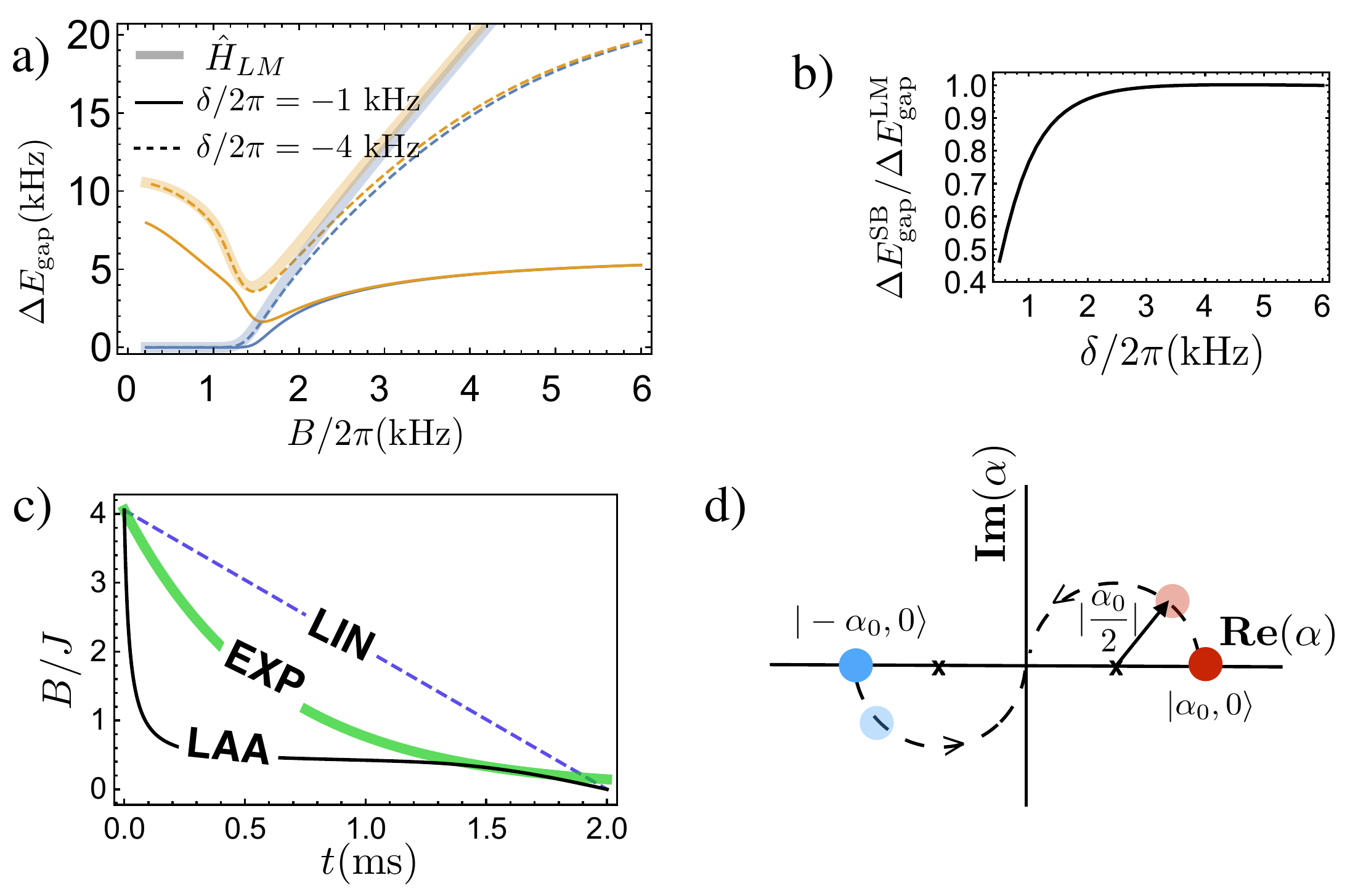}
 \caption{(a) The energy gap between the three lowest energy states of the $\hat H_{\rm LM}$ and $\hat{H}$, for two values of detuning $\delta$. Ramps were optimized with respect to the gap between the ground-state and the first excited state in the same parity sector which is shown in orange, and is referred to as the main gap throughout the paper. The gap between the ground-state and the lowest-excited state (of opposite parity), which below the critical point $B_c/(2\pi) \sim 1.5$~kHz is almost degenerate with the ground-state, is shown in blue. The presence of the resonance near the critical point modifies the position of the critical point and diminishes the magnitude of the main gap. (b) Scaling of the main gap of the spin-phonon Hamiltonian $\hat{H}$ with detuning, normalized with respect to the corresponding LM with fixed $J$. (c) The transverse field ramps used for the EXP (green),  LIN (dashed blue),  and LAA (black) protocols. Panels (a)-(c) use $N=20$ for calculations. (d) Schematic of the disentangling protocol to extract a pure spin cat-state from the spin-phonon ground-state $\vert \psi_{{\rm SB},\, I}^G \rangle$. At the end of the ramp, we quench the detuning $\delta \rightarrow 2\delta$ and evolve the system for an additional duration $t_d = \pi/|2\delta|$ at fixed $B=0$. The phonon states start at opposing coherent amplitudes and undergo a spin-dependent coherent displacement which maps them to the phonon vacuum state. See Sec.~\ref{sec:ramps} and Appendix A for further details.}
 \label{fig:gap}
\end{figure}

Figure~\ref{fig:gap} illustrates the two cases, which can be characterized by their energy spectra. In Panel (a) we plot example spectra for $\delta/ (2\pi) = -1$~kHz and $\delta/(2\pi)=-4$~kHz. In the former case, the detuning is chosen close to the critical field $B_c$, which leads to a large deviation in the spectrum near the critical point and related reduction in the minimum energy gap, while in the latter the spectrum near the critical point collapses to the effective LM result. Panel (b) examines the magnitude of the energy gap at the critical point as a function of detuning, relative to the LM.

\begin{figure*}[!]
 \includegraphics[width=0.85\textwidth]{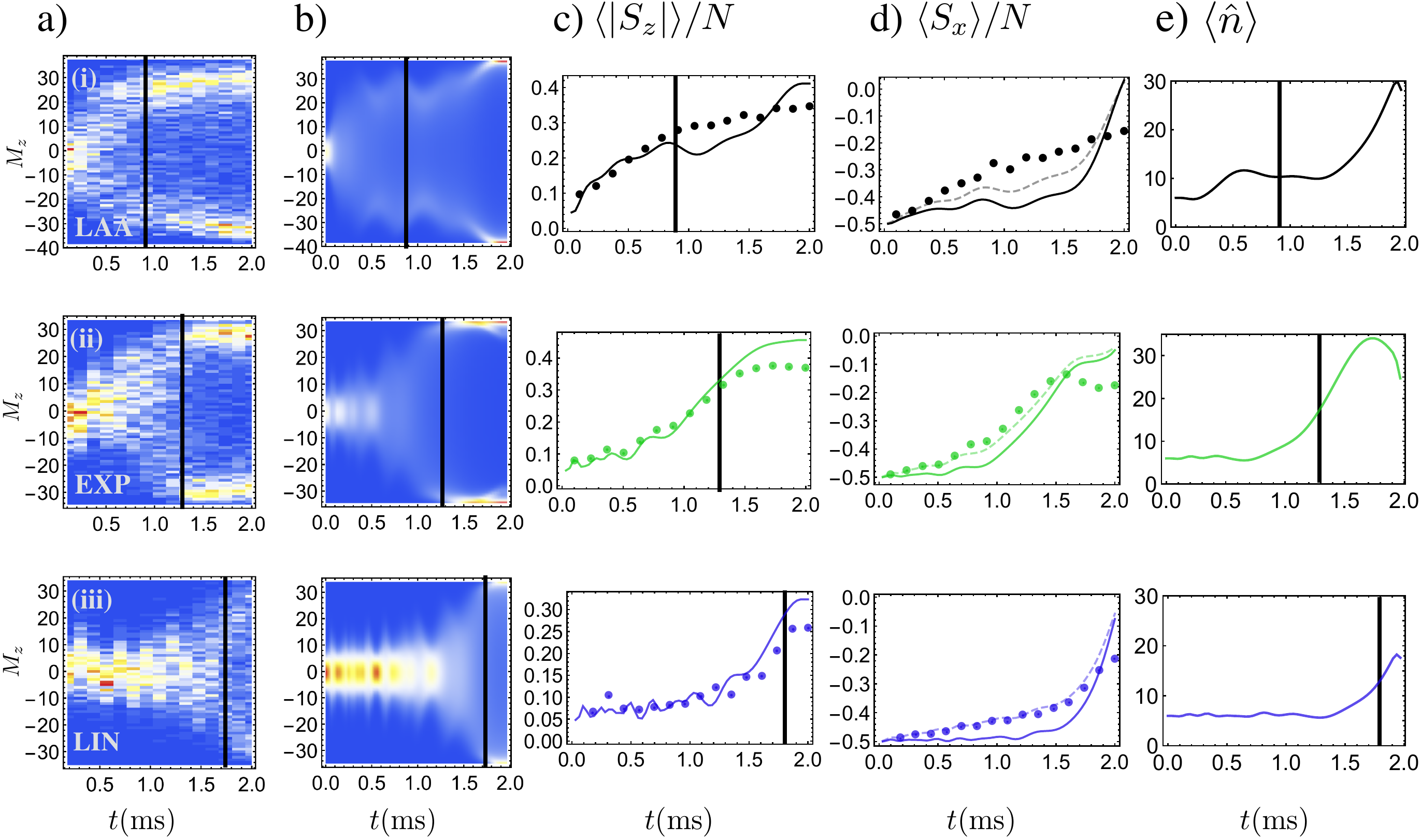}
 \caption{Comparison of experimental results  and theoretical (full spin-phonon) calculation for dynamics of spin observables. 
 We benchmark the $\hat S_z$ distribution functions extracted from experimental measurements of the spin-projection $M_z^{\rm exp}$ [column (a)] and theoretical simulations $M_z^{\rm th}$ [column (b)] at $\delta/(2\pi)= -1$~kHz with spin-dependent force chosen such that $J/(2\pi)=1.75$~kHz. Respective ion numbers are $N = 69$ [LIN -- row (i)], $N = 68$ [EXP -- row (ii)], $N = 76$ [LAA -- row (iii)]. We include a thermal phonon occupation $\bar n=6$ and neglect decoherence in these theoretical calculations. The double peak structure reflects the parity symmetry of the ground-state and is visible in both experiment and theory results.
 (c), (d), and (e) show the corresponding mean values of the magnetization $\langle |\hat{S}_z| \rangle$, spin-projection $\langle \hat{S}_x \rangle$, and mean phonon number $\langle \hat n \rangle $. The filled circles are experimental measurements, the solid and dashed lines are the theory results without and with dephasing $\Gamma_{el}=280$ s$^{-1}$. We indicate the time at which the critical point is reached in each ramp by a vertical line. We highlight that for the LAA ramp this must be interpreted with care, given that the majority of the ramp duration is spent in close proximity to the critical point [see Fig.~\ref{fig:gap}~(c)].  }
 \label{fig:ExpResults_ExpecValues}
\end{figure*}

\section{Protocols For Ground-state Preparation} \label{sec:ramps}

In this section, we consider different protocols to efficiently prepare the ground-state of the Dicke model for $B \rightarrow 0$. In all cases, we start with a large initial transverse field $B_i\equiv B(t=0)\gg |J|$, such that the spin-component of the ground-state is all spins polarized along $-\hat e_x$. For the experimental work, Doppler-limited cooling of the phonon degree of freedom means that the initial phonon state should be described by a thermal mixed state $\hat{\rho}_{\bar{n}}$ with mean occupation $\bar{n}=6$, such that the full spin-phonon state is characterised as $\hat{\rho}_{SB}(0) = \vert -N/2 \rangle_x \langle -N/2 \vert_x \otimes \hat{\rho}_{\bar n}$. We emphasize that starting with an initial thermal distribution does not change the physics discussed in the previous section (in which the phonon component of the ground-state in the normal phase was taken to be vacuum).  

If $B(t)$ is ramped to zero adiabatically, the ground-state component of the initial ensemble will follow the ground-state of the instantaneous Hamiltonian $\hat{H}(t)$, finishing in the 
state $\vert \psi_{{\rm SB}}(\tau)\rangle =\vert \psi_{{\rm SB}, I}^{GS}\rangle$, which is a spin-phonon cat-state [see Eq.~(\ref{eqn:SpinPhCat})]. On the other hand, for $B\rightarrow0$ the ground state of the LM is the spin cat-state 
$\vert \psi^{GS}_{{\rm S}, I} \rangle= \left(\vert N/2 \rangle_z + \vert -N/2 \rangle_z \right)/\sqrt{2}$. 

We note that the spin-phonon cat-state is an equivalent metrological resource to the spin cat-state in terms of the spin degree of freedom, e.g. it has the same sensitivity to rotations about $\hat{S}_z$.
However, it is difficult to experimentally probe features of the entangled spin-boson cat-state, such as the coherences. Specifically, while we can individually measure spin and phonon degrees of freedom \cite{Gilmore2017}, we lack the ability to simultaneously control and measure the phonon and spin degrees of freedom, which is crucial given that tracing out the phonon mode from the spin-boson cat-state exponentially suppresses the coherence between the spin states $\vert N/2\rangle_z$ and $\vert -N/2 \rangle_z$ (see Appendix~\ref{app:cat} for details), leading to an unentangled mixed spin state. Instead, given that we do possess the appropriate degree of control over the spin degree-of-freedom, we discuss how one can easily recover the spin-only cat-state by disentangling the spins and phonons in $\vert \psi_{{\rm SB}, I}^{GS}\rangle$. 
In Fig.~\ref{fig:gap}(d) we illustrate one such scheme where at the end of the ramp ($B \rightarrow 0$), we quench the detuning $\delta \rightarrow \delta^{\prime} = 2\delta$. As discussed in detail in Ref.~\cite{Wall2017}, in this regime the spin dynamics generated by $\hat{H}$ is equivalent to that of an effective Ising model on the spins, as well as a spin-dependent displacement for the bosons. The spin-component of $\vert \psi_{{\rm SB}, I}^{GS}\rangle$ is an eigenstate of the Ising model and it will not evolve any further. The evolution of the phonons for $t_d=\pi/\delta^\prime$ coherently displaces them by $\beta(t_d) = - g_0\sqrt{N}/(2|\delta|) \langle S_z\rangle$. This  brings the phonons back to a vacuum state such that $\vert\pm\alpha_0,0\rangle\vert\pm N/2\rangle_z \rightarrow \vert0\rangle\vert\pm N/2\rangle_z$, thus disentangling the phonons and spins. Tracing the phonons out of the disentangled state recovers a spin-only cat-state $\vert \psi_{{\rm SB}, I}^{GS}\rangle \rightarrow \left(\vert N/2 \rangle_z + \vert -N/2 \rangle_z \right)/\sqrt{2}$. A schematic of the protocol is illustrated in Fig.~\ref{fig:gap}(d) and details are presented  in Appendix A. 

Next, we investigate various ramp profiles to determine an optimal protocol given the interplay of experimentally realistic parameters and constraints with the varying degrees of adiabaticity of each protocol. An optimal ramp should balance the two to achieve the highest possible ground-state fidelity.
Specifically, we study three different ramp-down protocols over a fixed time $\tau_{\mathrm{ramp}}$ (see Fig.~\ref{fig:gap}~(c) for time-traces of the applied transverse field):

(i) Linear (LIN): $B(t)=B_0(1-t/\tau_{\mathrm{ramp}})$. While the LIN protocol is the simplest to implement, its efficiency is trivially limited by the relatively large speed at which the critical point is crossed.

(ii) Exponential (EXP): $B(t) = B_0e^{-t/\tau}$. This is a commonly used improvement to the LIN ramp, in which the field strength is ramped at a comparatively fast rate at short-times, thus enabling the velocity at the critical point to be significantly slower relative to the LIN ramp.

(iii) Local Adiabatic Approximation (LAA). 
The utility of the (LAA) is greatest when one has a very detailed understanding of the energy spectrum and critical point of the model \cite{Richerme2013,Freericks2015}. Specifically, the LAA ramp is constructed based on the energy gap $\Delta$ of the specific Hamiltonian studied.  For the LM and spin-phonon models the relevant gap is that to the lowest excited state of the same parity [orange line in Fig.~\ref{fig:gap}(a)]. The actual lowest energy excited state [blue line in Fig.~\ref{fig:gap}(a)] becomes degenerate for small $B \ll |J|$, however, due to the symmetry of the Hamiltonian these states remain uncoupled. The ramp is designed to rigorously optimize the 
instantaneous velocity of the transverse field $\dot{B}(t)$ based on the 
instantaneous gap $\Delta(t)$ \cite{Freericks2015}, such that diabatic excitations are created uniformly throughout the protocol. The ramp profile is given by solving $\dot{B}(t)  = \frac{1}{\gamma}\Delta(t)^2$ with 
\begin{equation}
\gamma  = \frac{\tau_{\mathrm{ramp}}}{\int_0^{B_0} dB \frac{1}{\Delta(B)^2}}
\end{equation}
a scaling factor determined from the initial and final magnetic field strengths and the total duration of the ramp.  From these  equations  we identify that the instantaneous ramp speed will be slowest when the gap $\Delta$ is minimal and fastest 
when $\Delta$ is largest.

\section{Experimental Results}\label{sec:expresults}


Resonant $124$~GHz microwave pulses generate arbitrary collective spin rotations, enabling preparation of the initial state $\left|-N/2\right\rangle_{x}$. Resonant microwaves also implement the effective transverse magnetic field $B(t)$ of Eq.~(\ref{eq:HField}) in the rotating frame of the qubit, where the initial field $B(t=0)/(2\pi) \approx 7.1$~kHz is ramped to zero with a voltage-controlled microwave attenuator. At the end of a ramp the magnetization in a particular projection-axis is measured by rotating that direction to the $\hat{z}$-axis and measuring the global ion fluorescence scattered from the Doppler cooling laser on the cycling transition for ions in $\left|\uparrow\right\rangle _{z}$ \cite{Martin2017_OTOC,Bohnet2016,Sawyer2012modeAndTempSpec,Biercuk2009}. We count the total number of photons collected on a photomultiplier tube in a detection period, typically $5$~ms. From the independently calibrated photons collected per ion, we can infer the state populations, $N_{\uparrow}$ and $N_{\downarrow}$, and calculate the spin-projection $M_z \equiv N_{\uparrow}-N/2$ for each experimental shot. From these measurements and repeated experimental trials we construct the full probability distribution $P(M_z)$.

Throughout the ramp, the spin-dependent ODF and detuning $\delta$ are constant, corresponding to fixed values of $J,\,g_0/\sqrt{N} \ll B(0)$. For the 2 ms ramp measurements of Fig. (3) where $N\approx70$, the effective spin-spin interaction and spin-mode coupling are $J/(2\pi) \approx 1.75$~kHz and $g_0/(2\pi) \approx 1.32$~kHz. 

Off-resonant light scattering from the detuned ODF laser beams is the dominant source of decoherence. For our experimental setup this decoherence is well described by single-particle dephasing with $\Gamma_{el} \approx 120 s^{-1}$ estimated from independent measurements at B=0 \cite{Uys2010}.

The exponential ramp time constant was set to $\tau\approx0.6$~ms. This value was chosen by experimentally optimizing the magnetization $\langle |\hat{S}_z| \rangle$ of the state at the end of the ramp protocol. In order to account for the experimental uncertainty in the spin-phonon coupling $g_0$ and consequently the spin-spin interaction $J$, we have optimized the LAA ramp in a similar manner, varying the estimated location of the critical point $B_c \sim J$ and optimizing via the magnetization $\langle |\hat{S}_z| \rangle$ of the final state. The exact functional form of the LAA ramp implemented in the experiment was determined by solving the associated ODE in Sec.~\ref{sec:ramps} with respect to the energy gap of the LM Hamiltonian. Further details can be found in Appendix.~\ref{app:RampOpt}. The profile of the transverse field for each ramp is illustrated  in Fig.~\ref{fig:gap}~(c).

In the experiment, the detuning $\delta$ is set to be close to the COM mode to generate the desired all-to-all coupling. Moreover, $\delta /(2\pi)=-1$~kHz is typically chosen to maximise the magnitude of the effective spin-spin Ising interaction relative to relevant sources of single-particle decoherence. For this condition the resonance lies close to the critical point of the related LM, as discussed previously and illustrated in Fig.~\ref{fig:gap}~(a). Hence, we expect the dynamics to deviate strongly from an effective spin-only description and that the phonons will play a crucial role in understanding the dynamics and efficiency of the ramp protocols. 


A complete characterization of the prepared state by measurement of the ground-state fidelity becomes exceedingly difficult for large systems. Instead, we characterize the state throughout the ramps using global spin observables. In particular, we focus on measurements of the magnetization $\vert \hat{S}_z \vert$ and the collective spin-projection along $\hat{S}_x$. In Fig.~\ref{fig:ExpResults_ExpecValues}, we show the complete distribution functions of the experimentally measured observables, as well as the mean-values. We compare the experimental data to theoretical calculations of the full spin-phonon model neglecting decoherence and assuming an initial thermal state for the phonons with mean occupation $\bar{n} = 6$. The dynamics of the Dicke model was solved using numerical integration of the Schr\"{o}dinger equation and thermal averaging of expectation values. We assumed the spin degree of freedom is restricted to the fully symmetric Dicke manifold ($S=N/2$) and used a truncated Fock basis for the phonon degree of freedom. Here, the appropriate truncation is estimated from the distribution of the final ground-state and is increased to check that results are converged.  

The experimental and theoretical spin distributions show good qualitative agreement. In particular, in both simulations and experimental data we see a clear transition to a bimodal structure in the spin distribution $P(M_z)$ of the spin-projection $M_z$ as the field strength is ramped down through the critical point (indicated by the black vertical line in each plot), consistent with the expected superposition of spins all up and all down as $B \rightarrow 0$ (although these results cannot be used to verify any coherence between these states). The ``smearing" out of the bimodal $P(M_z)$ distribution is largely attributable to the initial thermal distribution of phonons. 

A more direct comparison of the experimental data and theoretical calculations is made via the mean-values of the magnetization $\langle \vert \hat{S}_z \vert \rangle$ and spin-projection $\langle\hat{S}_x\rangle$, as shown in Figs.~\ref{fig:ExpResults_ExpecValues}~(c) and (d). We observe a depolarization along $S_x$ and increase in magnetization as the system crosses the critical point (again, indicated by the black vertical line in each plot) for all ramps. We observe good agreement between experimental measurements of the magnetization and our theoretical calculations, particularly for the EXP and LIN.  There is a larger deviation between experiment and theory for the mean spin-projection $\langle\hat{S}_x\rangle$ as, unlike the magnetization, this observable is strongly affected by dephasing mechanisms. However, the depolarization at large $B$, most clearly evidenced in the LIN protocol, is faster than expected for the estimated dephasing of $\Gamma_{el} \approx 120$~s$^{-1}$. Specifically, in order to account for this depolarization we instead need to include a larger dephasing of $\Gamma_{el} = 280$~s$^{-1}$. This dephasing could be a result of the experimental system going beyond the Lamb-Dicke regime, which is implicitly assumed in the derivation of the Dicke Hamiltonian Eq.~(\ref{eq:HI}) \cite{Cirac1995,Bohnet2016}. Since including the effects of decoherence, in combination with the large number of phonons and spins is computationally challenging, we crudely model the effect of dephasing by $\langle \hat{S}_x\rangle \rightarrow \langle \hat{S}_x \rangle e^{-\Gamma t}$ and $\langle \hat{S}_z\rangle\to \langle \hat{S}_z \rangle$ where $\Gamma=\Gamma_{el}/2$ \cite{Huelga1997}.  Strictly modelling dephasing in this manner is only valid in the $B=0$ limit, however, for $B \gg J$ or $B \ll J$ it is a reasonable approximation as in the former case the dynamics is dominated by single-particle physics, while in the latter the decoherence commutes with the spin-phonon coupling [Eq.~(\ref{eqn:SpinPhCoupling})]. The faint dashed lines in Fig.~\ref{fig:ExpResults_ExpecValues}(b) show the results with dephasing for $\langle \hat{S}_x \rangle$. With this simplistic treatment of decoherence, we highlight that $\langle |\hat{S}_z| \rangle$ remains unchanged as the decoherence commutes with $\hat{S}_z$, whereas in an exact treatment the depolarization of $\langle \hat{S}_x \rangle$ would consequently lead to a reduction in the final magnetization. While this approximation is crude, for the EXP and LIN ramps it improves the agreement between theory and experiment for $\langle \hat{S}_x \rangle$. However, it is insufficient to model the LAA ramp as the majority of the evolution is spent at $B\sim J$.


We generically find that the LAA and EXP ramps produce larger final magnetization [$\langle |\hat{S}^z| \rangle/N = 1/2$ ideally for $B(t) = 0$] than the simpler LIN ramp. This coarse observable indicates that the LAA and EXP ramps are preferred ramping protocols for adiabatic state preparation.

We also compute the expected mean phonon occupation $\langle \hat{n} \rangle$ for each of the ramps in Fig.~\ref{fig:ExpResults_ExpecValues}(e). We find for all ramps the occupation quickly becomes macroscopic below the critical point, and the final occupation is consistent with that of the superradiant ground state, $\langle \hat{n} \rangle \equiv \vert\alpha_0\vert^2 \sim 30$. While we did not measure this quantity experimentally, the phonon occupation is an accessible quantity \cite{Sawyer2012} and could be quantitatively verified in future experiments.

We further quantify the adiabaticity of the ramps on relevant experimental time-scales by performing ideal theory calculations of ground-state fidelity. We use the spin-phonon model neglecting decoherence and ignoring the initial thermal phonon occupation $\bar{n} = 0$. The fidelity to the spin-phonon cat-state state is then defined as $\mathcal{F}^{SB}_{GS} \equiv \vert \langle \psi(\tau_{\mathrm{ramp}}) \vert \psi^{GS}_{SB,I} \rangle \vert^2$.  We find only the LAA ramp has an appreciable fidelity $\mathcal{F}^{SB}_{GS} \simeq 0.25$ (in contrast, $\mathcal{F}^{SB}_{GS} < 10^{-3}$ for the EXP and LIN protocols). This poor outcome is attributable to modification of the energy spectrum due to the proximity of the resonant regime $B\sim |\delta|$ to the QCP at $B\sim J$ for the experimental detuning of $\delta/(2\pi)=-1$~kHz. Specifically, the significant reduction in the main energy gap relative to the LM implies much longer ramp durations are required to maintain adiabaticity.

However, we must point out that the ground-state fidelity in the experiment is further limited by a combination of single-particle dephasing and the thermal occupation  of the phonons. For an array of $N\approx70$ ions and current dephasing of $\Gamma\approx60$~s$^{-1}$, the off-diagonal coherences of the cat-state decay exponentially as $e^{-N\Gamma t} \sim 10^{-4}$ for $\tau_{\rm ramp} = 2$~ms. Assuming an adiabatic ramp such that the cat-state is prepared perfectly in the absence of decoherence, the fidelity to the cat-state with dephasing can then be crudely estimated as $\mathcal{F}^{SB}_{GS,\Gamma} \approx (1 + e^{-N\Gamma t})/2$. In addition to this, the achievable fidelity is also limited by the initial thermal occupation $\bar{n} \approx 6$ of the COM mode. This reduces the actual ground-state fraction of the initial spin-boson state by a factor of $1/(\bar n+1)$, and thus would reduce the achievable fidelity to the final ground-state by the same factor, $\mathcal{F}^{SB}_{GS,\bar{n}} \rightarrow \mathcal{F}^{SB}_{GS,\bar{n}=0}/(\bar{n}+1)$. The combination of dephasing and thermal phonon occupation leads us to the prediction $\mathcal{F}^{SB}_{GS} \lesssim (1 + e^{-N\Gamma t})/[2(\bar{n}+1)] \approx 0.07$. This fidelity is negligibly small, particularly in the context that we require $\mathcal{F}^{SB}_{GS} > 1/2$ to differentiate from a statistical mixture of the degenerate ground-states. 


\begin{figure}[!]
 \includegraphics[width=0.5\textwidth]{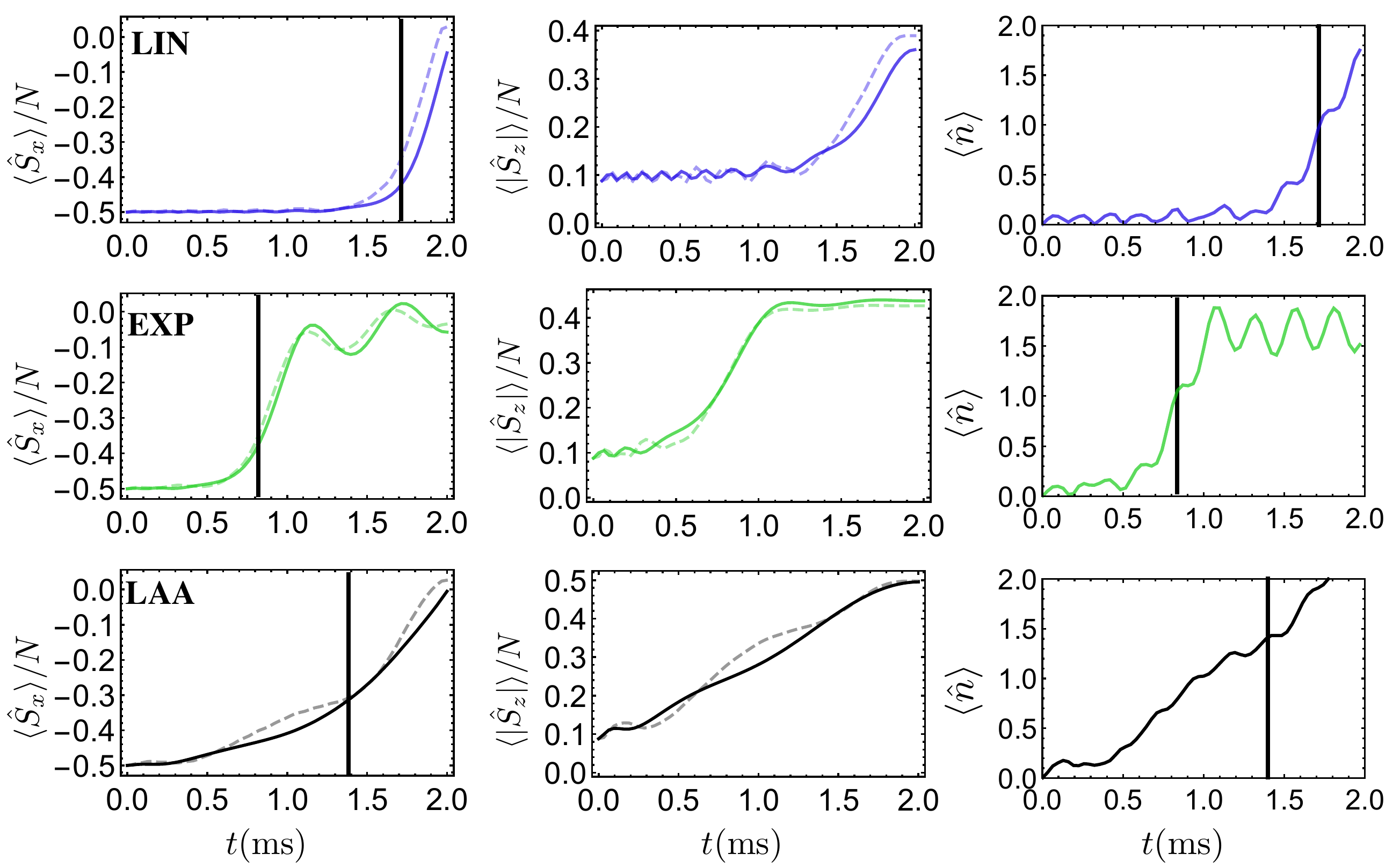}
 \caption{Dynamics for different ramp protocols at $\delta/(2\pi)= -4$~kHz. (a) and (b) show a comparison of dynamics of the spin observables generated by the LM (solid) with the full spin-boson (dashed). (c) The mean phonon number for the Dicke model.  The parameters are adjusted to keep the same effective spin-spin interactions $J/(2\pi)=1.75$~kHz, $N=20$ and initial $\bar n=0$. }
\label{fig:IdealResults_ExpecValues}
\end{figure}

\section{Theoretical predictions for next generation of experiment}

The detailed analysis of experimental results presented above allows us to devise methods which can lead to improved ground-state fidelities.  
\begin{figure}[!]
 \includegraphics[width=0.5\textwidth]{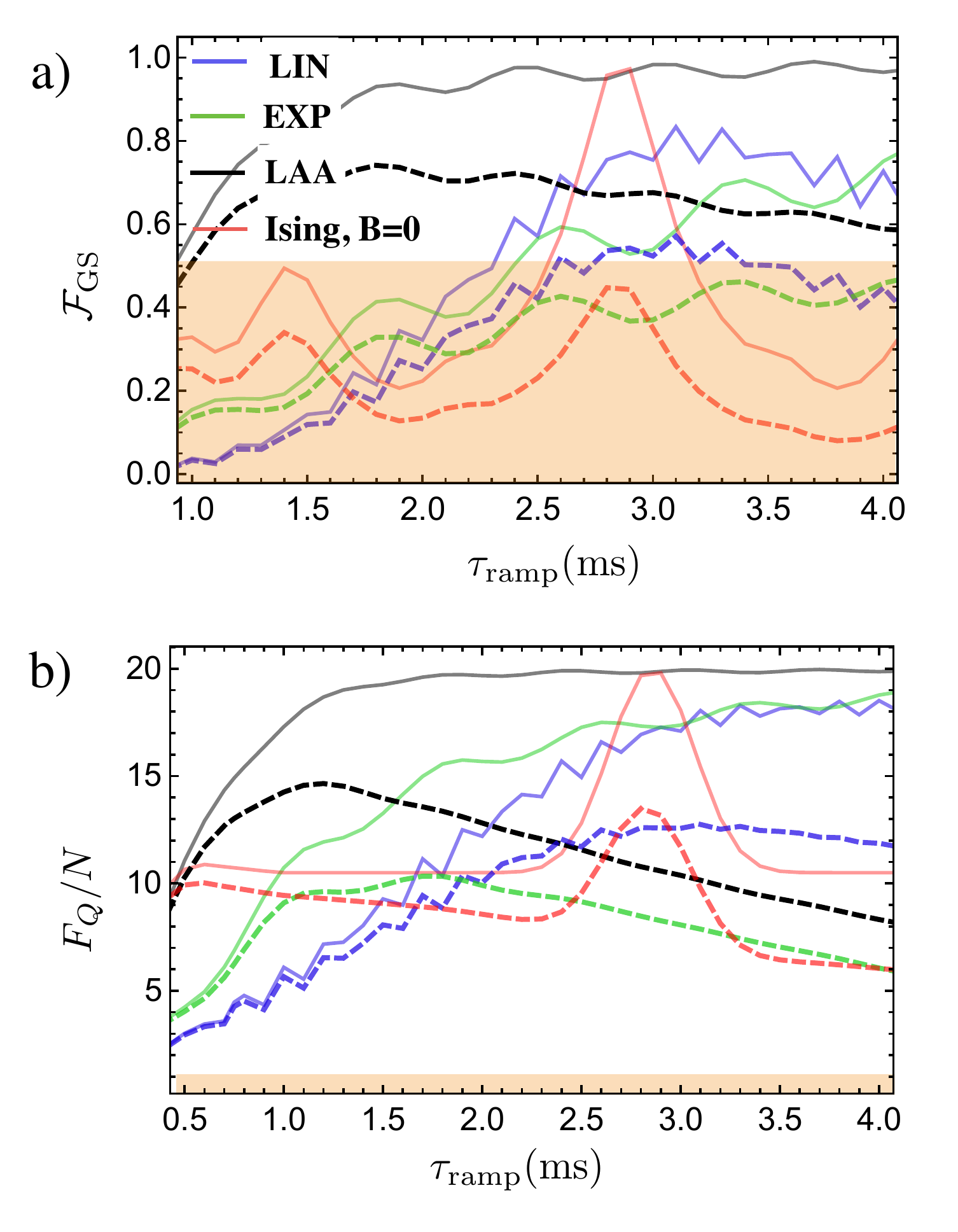}
\caption{Ground state fidelity to the spin-phonon cat-state $\mathcal F^{SB}_{GS}$ (Dicke model) and spin cat-state $\mathcal F^{S}_{GS}$ (Lipkin model), and QFI per particle $F_Q/N$ as a function of ramp duration, $\tau_{\rm ramp}$ for $N=20$, $\bar n=0$, and $\delta/(2\pi)= -4$~kHz. The solid lines correspond to the ideal case $\Gamma = 0$, while the dashed lines include the effect of dephasing $\Gamma=12$~s$^{-1}$ on the ground-state fidelity and QFI for the effective LM. Here, we account for all relevant decoherence processes and define $\Gamma = (\Gamma_{el} + \Gamma_{du} + \Gamma_{ud})/2$ as the total decoherence rate. Here $\Gamma_{ud(du)}$ is the rate of spontaneous emission (absorption) and is assumed to be negligible compared to dephasing ($\Gamma_{el} = 20$~s$^{-1}$, $\Gamma_{du} = \Gamma_{ud} = 2$~s$^{-1}$). 
 In (a) we use the shaded region to indicate an approximate confidence bound for `catness' of the final state. The shaded region in (b) indicates the loss of any metrologically useful entanglement.}
 \label{fig:N20Ideal}
\end{figure}
The first modification to the experimental procedure is to avoid choosing a detuning close to the critical point, which eliminates the complications arising from undesirable spin-phonon entanglement, specifically, the relative reduction in the energy gap. To this end, we set the detuning to $\delta/(2\pi)=-4$~kHz while keeping the effective spin-spin interaction coefficient $J$ of the LM model constant, thus realizing the same effective spin model. By shifting the detuning away from the critical field strength the main gap at the QCP is significantly increased and thus reduces the creation of non-adiabatic excitations for typical ramping times. 

The next generation of experiments are expected to implement electromagnetically induced transparency (EIT) cooling schemes \cite{Roos2016}. This will substantially reduce the initial thermal phonon occupation to $\bar n\simeq0.2$ which corresponds to an initial ground-state fraction (of the full spin-phonon Hamiltonian) of $>80\%$, greatly enhancing the achievable spin-phonon cat-state fidelity [$\mathcal{F}^{SB}_{GS}\propto 1/(\bar{n} + 1$)]. 

With this improved initial state preparation, the remaining barrier to achieving appreciable ground-state fidelities for large arrays of ions remains the dephasing $\Gamma$. Improved cooling will aid in eliminating the spurious sources of decoherence, particularly those due to going beyond the Lamb-Dicke regime. Beyond this, we then anticipate a $4$-fold reduction of the spontaneous emission relative to coherent interaction strength by increasing the angle between the ODF beams by a factor of $2$ \cite{Bohnet2016}. The resulting reduction in decoherence $\Gamma\sim 15$~s$^{-1}$ will allow  us to reach fidelities of $\mathcal{F}^{SB}_{GS} \sim 0.4-0.8$ for a crystal of 20 ions under a ramp of  $\tau_{\mathrm{ramp}} \sim 1-4$ms. Beyond our specific system, there exist alternative trapped ion setups with more favorable rates of decoherence, such as Yb$^+$ or Ca$^+$ ions in a linear RF trap \cite{Monroe2014,Zhang2017, Blatt2008, PhysRevLett.106.130506}.

Figure~\ref{fig:IdealResults_ExpecValues} shows our theoretical predictions for spin-observables for the LIN, EXP and LAA ramps with duration $\tau_{\mathrm{ramp}} = 2$ms, using the modified parameters discussed here and neglecting dephasing $\Gamma = 0$. 
By removing the resonance away from the QCP, we immediately observe the dynamics of the LM and the Dicke model (with initial $\bar{n} = 0$) can display good agreement. 

We also calculate the fidelity to the ground-state for the same parameters as a function of ramp duration in Fig.~\ref{fig:N20Ideal}. Neglecting thermal phonon population and decoherence, we find that significant fidelities, $\mathcal{F}^{SB}_{GS} > 1/2$, can be achieved with ramps of duration $\tau_{\mathrm{ramp}} \simeq 2-4$ms depending on the chosen protocol. Given that the spin-phonon dynamics now closely track the spin-dynamics in this parameter regime, we estimate the effects of decoherence by considering only the spin-dynamics under the LM with $\Gamma = 12$~s$^{-1}$ ($\Gamma_{el} = 20$~s$^{-1}$, $\Gamma_{du} = \Gamma_{ud} = 2$~s$^{-1}$). For these parameters, we estimate a maximum fidelity of $\mathcal{F}^{SB}_{GS} \approx 0.75$ for the LAA ramp or $\mathcal{F}^{SB}_{GS} \approx 0.5$ for the EXP and LIN ramps.

While achieving ground-state fidelities of $O(1)$ with large number of ions in the presence of decoherence is very challenging, one may also use ramping protocols to generate highly entangled states. 
We use the quantum Fisher information (QFI) as a witness to characterize the depth of entanglement of the state as a function of ramp time. For an arbitrary quantum state the QFI is defined as \cite{Braunstein1994}:
\begin{equation}
 \mathcal{F}_Q = 2\sum_{k,l} \frac{\left(\lambda_k - \lambda_l\right)^2}{\lambda_k + \lambda_l} \vert \langle \lambda_k \vert \hat{G} \vert \lambda_l \rangle \vert^2
\end{equation}
where $\hat{G}$ is a chosen generator, $\lambda_{k,l}$ the eigenvalues of the density matrix and $|\lambda_{k,l}\rangle$ the corresponding eigenvectors. 
We optimize over all generators $\hat{G}= \vec{v}\cdot\hat{\vec{S}}$ which correspond to collective rotations about an axis of the Bloch sphere along the unit vector $\vec{v} = [\mathrm{sin}(\phi)\mathrm{cos}(\theta), \mathrm{sin}(\phi)\mathrm{sin}(\theta),\mathrm{cos}(\phi)]$ parameterized by $(\theta, \phi)$, which are the polar and azimuthal angles, respectively. For the ground-state of the Dicke Hamiltonian and LM the optimal rotation axis will be $\hat{G} = \hat{S}_z$, and this is used in Fig.~\ref{fig:N20Ideal}(b). For the Dicke model, we calculate the QFI of the final state without any disentangling operation to remove the phonons. 

In order to benchmark the utility of the ramp it is useful to compare our results to the QFI of the state generated by the Ising dynamics. This is motivated by previous experimental and theoretical results by several co-authors discussed in Ref.~\cite{Bohnet2016}. 
Here, we perform theoretical simulations of 20 ions, using the LM and the Dicke model, for the three different ramping protocols. In Fig.~\ref{fig:N20Ideal}, we compare the results of these state preparation protocols to the dynamics of the QFI under the effective all-to-all Ising Hamiltonian given by Eq.~(\ref{eq:HTFIM}) with $B=0$, and the Dicke Hamiltonian with $B=0$, fixing the detuning $\delta/(2\pi) =-4$~kHz and $J$ to be the same as for the adiabatic protocol. In the case of the all-to-all Ising Hamiltonian, we prepare an initial coherent spin-state pointing along $\hat e_x$, identical to the large $B$ state for the ramping protocols. The peak in the QFI at $\sim 3$~ms corresponds to the spin cat-state for the Ising dynamics, which is a superposition of $\left(|N/2\rangle_x + |-N/2\rangle_x\right)/\sqrt{2}$, in contrast to the adiabatically prepared spin-phonon cat-state for which the spins are aligned along $\pm \hat{e}_z$. 

In the absence of decoherence (solid lines), a LAA ramp of $\tau_{\mathrm{ramp}} \sim 1$~ms matches the maximal entanglement generated by the Ising Hamiltonian three times faster. Thus while preparing the true ground-state is generically difficult, one can still generate a useful bimodal distribution along $P(M_z)$ which as the QFI demonstrates,  is due to a coherent superposition rather than a statistical mixture. This implies that one can produce spectroscopically useful entangled states using non-adiabatic ramps.

We find that in the presence of decoherence, specifically dephasing and spontanenous emission, the ability to generate entanglement rapidly with a ramping protocol does have some benefit in comparison to the Ising Hamiltonian. This is evidenced by the LAA ramp, which is able to generate a slightly larger Fisher information $F_Q$ at $\tau_{\mathrm{ramp}}\sim 1$~ms than that achievable by the Ising Hamiltonian at $\tau_{\mathrm{ramp}}\sim 3$~ms.


While the ramping protocols provide a procedure to rapidly produce entanglement, the generated states are not neccessarily amenable to simple inteferometric schemes, such as Ramsey interferometry, as compared to, e.g., the squeezed states produced by the Ising Hamiltonian. Generically, cat-like states require single-particle-resolved detection to fully unlock their metrological usefulness \cite{Boixo_2007}, and this should be possible with the implementation of the single-site resolution capability of the experimental setup. Beyond this, however, there are also proposals for time-reversal schemes \cite{Macri_2016,Davis_2016,Linnemann_2016,Hosten_2016}, intrinsically related to recent measurements of out-of-time-order correlations \cite{Martin2017_OTOC,Meier_2017}.

\section{Summary and Conclusions}

In this work, we have benchmarked the behavior of a 2D ion crystal of $\sim 70$ ions in the presence of a transverse magnetic field. The Hamiltonian dynamics result from the interplay of the coupling between the transverse field and the spins $\hat H_B$, the phonons $\hat H_{\omega; I}$, and the spin-phonon coupling $\hat H_{\rm SB; I}$. We implemented different ramps, transforming an approximately prepared ground-state of the normal phase to the ground-state of the superradiant phase of the Dicke Hamiltonian. Whilst the ground-state physics of the Dicke Hamiltonian mirrors that of the related LM, we demonstrated that the energy spectrum can deviate significantly and has important implications for adiabatic state preparation in a trapped ion quantum simulator. 


We have demonstrated that the appearance of states with large magnetization $\langle|\hat{S}_z|\rangle$ and a bimodal structure in the related distribution function $P(M_z)$, consistent with the theoretical predictions. Our theoretical calculation of the full spin-phonon model displays qualitative agreement, particularly for the observed spin distribution functions.

We also presented theoretical results demonstrating that by detuning further away from the COM mode allows one to modify the level structure of the Dicke Hamiltonian in such a way that it mimics that of the LM. However, we note that the detuning is still small enough that we can maintain the same spin-spin couplings with a modest increase in laser power, or simple rearrangements to the ODF laser beam configuration. However, the detuning still places us in a regime where the full spin-boson dynamics should be considered. Together with a modest reduction in ion number, decoherence rate and implementation of EIT cooling of the motional modes of the crystal, we predict that the next generation of experiments could prepare the true $B=0$ ground-state (cat-state) for a mesoscopic number of ions with a fidelity of $\mathcal{F}^{SB}_{GS}=0.5 - 0.75$.

Finally we have shown that ramp protocols can be a useful method for generating metrologically useful entangled states. In particular, we predict the generation of large Fisher information per particle ($\gg1$) even for very short (non-adiabatic) ramps, outperforming the equivalent entanglement generated by application of the generic squeezing Hamiltonian. The rapid generation of multipartite entanglement is a key tool in combating the fragility of such states to decoherence. 


\acknowledgements
The authors acknowledge fruitful discussions with J. Marino, M. Holland and K. Lehnert. 
A.~M.~R acknowledges support from Defense Advanced Research Projects Agency (DARPA) and Army Research Office grant W911NF-16-1-0576, NSF grant PHY1521080, JILA-NSF grant PFC-173400, and the Air Force Office of Scientific Research and its Multidisciplinary University Research Initiative grant FA9550-13-1-0086. M.G. acknowledges support from the DFG Collaborative Research Center SFB1225 (ISOQUANT). E. J. also acknowledges support from Leopoldina Fellowship Programme. JKF and JC acknowledge support from NSF grant PHYS-1620555. In addition, JKF acknowledges support from the McDevitt bequest at Georgetown. Financial support from NIST is also acknowledged.

\vspace{2 cm}
\bibliography{library}

\begin{thebibliography}{55}%
\makeatletter
\providecommand \@ifxundefined [1]{%
 \@ifx{#1\undefined}
}%
\providecommand \@ifnum [1]{%
 \ifnum #1\expandafter \@firstoftwo
 \else \expandafter \@secondoftwo
 \fi
}%
\providecommand \@ifx [1]{%
 \ifx #1\expandafter \@firstoftwo
 \else \expandafter \@secondoftwo
 \fi
}%
\providecommand \natexlab [1]{#1}%
\providecommand \enquote  [1]{``#1''}%
\providecommand \bibnamefont  [1]{#1}%
\providecommand \bibfnamefont [1]{#1}%
\providecommand \citenamefont [1]{#1}%
\providecommand \href@noop [0]{\@secondoftwo}%
\providecommand \href [0]{\begingroup \@sanitize@url \@href}%
\providecommand \@href[1]{\@@startlink{#1}\@@href}%
\providecommand \@@href[1]{\endgroup#1\@@endlink}%
\providecommand \@sanitize@url [0]{\catcode `\\12\catcode `\$12\catcode
  `\&12\catcode `\#12\catcode `\^12\catcode `\_12\catcode `\%12\relax}%
\providecommand \@@startlink[1]{}%
\providecommand \@@endlink[0]{}%
\providecommand \url  [0]{\begingroup\@sanitize@url \@url }%
\providecommand \@url [1]{\endgroup\@href {#1}{\urlprefix }}%
\providecommand \urlprefix  [0]{URL }%
\providecommand \Eprint [0]{\href }%
\providecommand \doibase [0]{http://dx.doi.org/}%
\providecommand \selectlanguage [0]{\@gobble}%
\providecommand \bibinfo  [0]{\@secondoftwo}%
\providecommand \bibfield  [0]{\@secondoftwo}%
\providecommand \translation [1]{[#1]}%
\providecommand \BibitemOpen [0]{}%
\providecommand \bibitemStop [0]{}%
\providecommand \bibitemNoStop [0]{.\EOS\space}%
\providecommand \EOS [0]{\spacefactor3000\relax}%
\providecommand \BibitemShut  [1]{\csname bibitem#1\endcsname}%
\let\auto@bib@innerbib\@empty
\bibitem [{\citenamefont {Leibfried}\ \emph {et~al.}(2005)\citenamefont
  {Leibfried}, \citenamefont {Knill}, \citenamefont {Seidelin}, \citenamefont
  {Britton}, \citenamefont {Blakestad}, \citenamefont {Chiaverini},
  \citenamefont {Hume}, \citenamefont {Itano}, \citenamefont {Jost},
  \citenamefont {Langer}, \citenamefont {Ozeri}, \citenamefont {Reichle},\ and\
  \citenamefont {Wineland}}]{Leibfried2005}%
  \BibitemOpen
  \bibfield  {author} {\bibinfo {author} {\bibfnamefont {D.}~\bibnamefont
  {Leibfried}}, \bibinfo {author} {\bibfnamefont {E.}~\bibnamefont {Knill}},
  \bibinfo {author} {\bibfnamefont {S.}~\bibnamefont {Seidelin}}, \bibinfo
  {author} {\bibfnamefont {J.}~\bibnamefont {Britton}}, \bibinfo {author}
  {\bibfnamefont {R.~B.}\ \bibnamefont {Blakestad}}, \bibinfo {author}
  {\bibfnamefont {J.}~\bibnamefont {Chiaverini}}, \bibinfo {author}
  {\bibfnamefont {D.~B.}\ \bibnamefont {Hume}}, \bibinfo {author}
  {\bibfnamefont {W.~M.}\ \bibnamefont {Itano}}, \bibinfo {author}
  {\bibfnamefont {J.~D.}\ \bibnamefont {Jost}}, \bibinfo {author}
  {\bibfnamefont {C.}~\bibnamefont {Langer}}, \bibinfo {author} {\bibfnamefont
  {R.}~\bibnamefont {Ozeri}}, \bibinfo {author} {\bibfnamefont
  {R.}~\bibnamefont {Reichle}}, \ and\ \bibinfo {author} {\bibfnamefont
  {D.~J.}\ \bibnamefont {Wineland}},\ }\href {\doibase 10.1038/nature04251}
  {\bibfield  {journal} {\bibinfo  {journal} {Nature}\ }\textbf {\bibinfo
  {volume} {438}},\ \bibinfo {pages} {639} (\bibinfo {year}
  {2005})}\BibitemShut {NoStop}%
\bibitem [{\citenamefont {Gilchrist}\ \emph {et~al.}(2004)\citenamefont
  {Gilchrist}, \citenamefont {Nemoto}, \citenamefont {Munro}, \citenamefont
  {Ralph}, \citenamefont {Glancy}, \citenamefont {Braunstein},\ and\
  \citenamefont {Milburn}}]{Gilchrist2004}%
  \BibitemOpen
  \bibfield  {author} {\bibinfo {author} {\bibfnamefont {A.}~\bibnamefont
  {Gilchrist}}, \bibinfo {author} {\bibfnamefont {K.}~\bibnamefont {Nemoto}},
  \bibinfo {author} {\bibfnamefont {W.~J.}\ \bibnamefont {Munro}}, \bibinfo
  {author} {\bibfnamefont {T.~C.}\ \bibnamefont {Ralph}}, \bibinfo {author}
  {\bibfnamefont {S.}~\bibnamefont {Glancy}}, \bibinfo {author} {\bibfnamefont
  {S.~L.}\ \bibnamefont {Braunstein}}, \ and\ \bibinfo {author} {\bibfnamefont
  {G.~J.}\ \bibnamefont {Milburn}},\ }\href@noop {} {\bibfield  {journal}
  {\bibinfo  {journal} {J. Opt. B: Quantum and Semiclassical Optics}\ }\textbf
  {\bibinfo {volume} {6}},\ \bibinfo {pages} {S828} (\bibinfo {year}
  {2004})}\BibitemShut {NoStop}%
\bibitem [{\citenamefont {Giovannetti}\ \emph {et~al.}(2004)\citenamefont
  {Giovannetti}, \citenamefont {Lloyd},\ and\ \citenamefont
  {Maccone}}]{Giovannetti1330}%
  \BibitemOpen
  \bibfield  {author} {\bibinfo {author} {\bibfnamefont {V.}~\bibnamefont
  {Giovannetti}}, \bibinfo {author} {\bibfnamefont {S.}~\bibnamefont {Lloyd}},
  \ and\ \bibinfo {author} {\bibfnamefont {L.}~\bibnamefont {Maccone}},\ }\href
  {\doibase 10.1126/science.1104149} {\bibfield  {journal} {\bibinfo  {journal}
  {Science}\ }\textbf {\bibinfo {volume} {306}},\ \bibinfo {pages} {1330}
  (\bibinfo {year} {2004})}\BibitemShut {NoStop}%
\bibitem [{\citenamefont {Strobel}\ \emph {et~al.}(2014)\citenamefont
  {Strobel}, \citenamefont {Muessel}, \citenamefont {Linnemann}, \citenamefont
  {Zibold}, \citenamefont {Hume}, \citenamefont {Pezz{\`e}}, \citenamefont
  {Smerzi},\ and\ \citenamefont {Oberthaler}}]{Strobel424}%
  \BibitemOpen
  \bibfield  {author} {\bibinfo {author} {\bibfnamefont {H.}~\bibnamefont
  {Strobel}}, \bibinfo {author} {\bibfnamefont {W.}~\bibnamefont {Muessel}},
  \bibinfo {author} {\bibfnamefont {D.}~\bibnamefont {Linnemann}}, \bibinfo
  {author} {\bibfnamefont {T.}~\bibnamefont {Zibold}}, \bibinfo {author}
  {\bibfnamefont {D.~B.}\ \bibnamefont {Hume}}, \bibinfo {author}
  {\bibfnamefont {L.}~\bibnamefont {Pezz{\`e}}}, \bibinfo {author}
  {\bibfnamefont {A.}~\bibnamefont {Smerzi}}, \ and\ \bibinfo {author}
  {\bibfnamefont {M.~K.}\ \bibnamefont {Oberthaler}},\ }\href {\doibase
  10.1126/science.1250147} {\bibfield  {journal} {\bibinfo  {journal}
  {Science}\ }\textbf {\bibinfo {volume} {345}},\ \bibinfo {pages} {424}
  (\bibinfo {year} {2014})}\BibitemShut {NoStop}%
\bibitem [{\citenamefont {Nielsen}(2006)}]{NIELSEN2006147}%
  \BibitemOpen
  \bibfield  {author} {\bibinfo {author} {\bibfnamefont {M.~A.}\ \bibnamefont
  {Nielsen}},\ }\href {\doibase
  http://dx.doi.org/10.1016/S0034-4877(06)80014-5} {\bibfield  {journal}
  {\bibinfo  {journal} {Rep. Math. Phys.}\ }\textbf {\bibinfo {volume} {57}},\
  \bibinfo {pages} {147 } (\bibinfo {year} {2006})}\BibitemShut {NoStop}%
\bibitem [{\citenamefont {Leroux}\ \emph {et~al.}(2010)\citenamefont {Leroux},
  \citenamefont {Schleier-Smith},\ and\ \citenamefont
  {Vuleti{\'{c}}}}]{Leroux2010c}%
  \BibitemOpen
  \bibfield  {author} {\bibinfo {author} {\bibfnamefont {I.~D.}\ \bibnamefont
  {Leroux}}, \bibinfo {author} {\bibfnamefont {M.~H.}\ \bibnamefont
  {Schleier-Smith}}, \ and\ \bibinfo {author} {\bibfnamefont {V.}~\bibnamefont
  {Vuleti{\'{c}}}},\ }\href {\doibase 10.1103/PhysRevLett.104.073602}
  {\bibfield  {journal} {\bibinfo  {journal} {Phys. Rev. Lett.}\ }\textbf
  {\bibinfo {volume} {104}},\ \bibinfo {pages} {073602} (\bibinfo {year}
  {2010})}\BibitemShut {NoStop}%
\bibitem [{\citenamefont {Ritsch}\ \emph {et~al.}(2013)\citenamefont {Ritsch},
  \citenamefont {Domokos}, \citenamefont {Brennecke},\ and\ \citenamefont
  {Esslinger}}]{Ritsch2013}%
  \BibitemOpen
  \bibfield  {author} {\bibinfo {author} {\bibfnamefont {H.}~\bibnamefont
  {Ritsch}}, \bibinfo {author} {\bibfnamefont {P.}~\bibnamefont {Domokos}},
  \bibinfo {author} {\bibfnamefont {F.}~\bibnamefont {Brennecke}}, \ and\
  \bibinfo {author} {\bibfnamefont {T.}~\bibnamefont {Esslinger}},\ }\href
  {\doibase 10.1103/RevModPhys.85.553} {\bibfield  {journal} {\bibinfo
  {journal} {Rev. Mod. Phys.}\ }\textbf {\bibinfo {volume} {85}},\ \bibinfo
  {pages} {553} (\bibinfo {year} {2013})}\BibitemShut {NoStop}%
\bibitem [{\citenamefont {Porras}\ and\ \citenamefont
  {Cirac}(2004)}]{Porras2004}%
  \BibitemOpen
  \bibfield  {author} {\bibinfo {author} {\bibfnamefont {D.}~\bibnamefont
  {Porras}}\ and\ \bibinfo {author} {\bibfnamefont {J.~I.}\ \bibnamefont
  {Cirac}},\ }\href {\doibase 10.1103/PhysRevLett.92.207901} {\bibfield
  {journal} {\bibinfo  {journal} {Phys. Rev. Lett.}\ }\textbf {\bibinfo
  {volume} {92}},\ \bibinfo {pages} {207901} (\bibinfo {year} {2004})},\
  \Eprint {http://arxiv.org/abs/0401102} {arXiv:0401102} \BibitemShut {NoStop}%
\bibitem [{\citenamefont {Kim}\ \emph {et~al.}(2009)\citenamefont {Kim},
  \citenamefont {Chang}, \citenamefont {Islam}, \citenamefont {Korenblit},
  \citenamefont {Duan},\ and\ \citenamefont {Monroe}}]{Kim2009}%
  \BibitemOpen
  \bibfield  {author} {\bibinfo {author} {\bibfnamefont {K.}~\bibnamefont
  {Kim}}, \bibinfo {author} {\bibfnamefont {M.-S.}\ \bibnamefont {Chang}},
  \bibinfo {author} {\bibfnamefont {R.}~\bibnamefont {Islam}}, \bibinfo
  {author} {\bibfnamefont {S.}~\bibnamefont {Korenblit}}, \bibinfo {author}
  {\bibfnamefont {L.-M.}\ \bibnamefont {Duan}}, \ and\ \bibinfo {author}
  {\bibfnamefont {C.}~\bibnamefont {Monroe}},\ }\href {\doibase
  10.1103/PhysRevLett.103.120502} {\bibfield  {journal} {\bibinfo  {journal}
  {Phys. Rev. Lett.}\ }\textbf {\bibinfo {volume} {103}},\ \bibinfo {pages}
  {120502} (\bibinfo {year} {2009})}\BibitemShut {NoStop}%
\bibitem [{\citenamefont {Zhang}\ \emph {et~al.}()\citenamefont {Zhang},
  \citenamefont {Pagano}, \citenamefont {Hess}, \citenamefont {Kyprianidis},
  \citenamefont {Becker}, \citenamefont {Kaplan}, \citenamefont {Gorshkov},
  \citenamefont {Gong},\ and\ \citenamefont {Monroe}}]{Zhang2017}%
  \BibitemOpen
  \bibfield  {author} {\bibinfo {author} {\bibfnamefont {J.}~\bibnamefont
  {Zhang}}, \bibinfo {author} {\bibfnamefont {G.}~\bibnamefont {Pagano}},
  \bibinfo {author} {\bibfnamefont {P.}~\bibnamefont {Hess}}, \bibinfo {author}
  {\bibfnamefont {A.}~\bibnamefont {Kyprianidis}}, \bibinfo {author}
  {\bibfnamefont {P.}~\bibnamefont {Becker}}, \bibinfo {author} {\bibfnamefont
  {H.}~\bibnamefont {Kaplan}}, \bibinfo {author} {\bibfnamefont
  {A.}~\bibnamefont {Gorshkov}}, \bibinfo {author} {\bibfnamefont {Z.-X.}\
  \bibnamefont {Gong}}, \ and\ \bibinfo {author} {\bibfnamefont
  {C.}~\bibnamefont {Monroe}},\ }\href@noop {} {\enquote {\bibinfo {title}
  {Observation of a many-body dynamical phase transition with a 53-qubit
  quantum simulator},}\ }\Eprint {http://arxiv.org/abs/arXiv:1708.01044}
  {arXiv:1708.01044} \BibitemShut {NoStop}%
\bibitem [{\citenamefont {Richerme}\ \emph
  {et~al.}(2013{\natexlab{a}})\citenamefont {Richerme}, \citenamefont {Senko},
  \citenamefont {Korenblit}, \citenamefont {Smith}, \citenamefont {Lee},
  \citenamefont {Islam}, \citenamefont {Campbell},\ and\ \citenamefont
  {Monroe}}]{Richerme2013}%
  \BibitemOpen
  \bibfield  {author} {\bibinfo {author} {\bibfnamefont {P.}~\bibnamefont
  {Richerme}}, \bibinfo {author} {\bibfnamefont {C.}~\bibnamefont {Senko}},
  \bibinfo {author} {\bibfnamefont {S.}~\bibnamefont {Korenblit}}, \bibinfo
  {author} {\bibfnamefont {J.}~\bibnamefont {Smith}}, \bibinfo {author}
  {\bibfnamefont {A.}~\bibnamefont {Lee}}, \bibinfo {author} {\bibfnamefont
  {R.}~\bibnamefont {Islam}}, \bibinfo {author} {\bibfnamefont {W.~C.}\
  \bibnamefont {Campbell}}, \ and\ \bibinfo {author} {\bibfnamefont
  {C.}~\bibnamefont {Monroe}},\ }\href {\doibase
  10.1103/PhysRevLett.111.100506} {\bibfield  {journal} {\bibinfo  {journal}
  {Phys. Rev. Lett.}\ }\textbf {\bibinfo {volume} {111}},\ \bibinfo {pages} {1}
  (\bibinfo {year} {2013}{\natexlab{a}})},\ \Eprint
  {http://arxiv.org/abs/1303.6983} {arXiv:1303.6983} \BibitemShut {NoStop}%
\bibitem [{\citenamefont {Richerme}\ \emph
  {et~al.}(2013{\natexlab{b}})\citenamefont {Richerme}, \citenamefont {Senko},
  \citenamefont {Smith}, \citenamefont {Lee}, \citenamefont {Korenblit},\ and\
  \citenamefont {Monroe}}]{Richerme2013_PRA}%
  \BibitemOpen
  \bibfield  {author} {\bibinfo {author} {\bibfnamefont {P.}~\bibnamefont
  {Richerme}}, \bibinfo {author} {\bibfnamefont {C.}~\bibnamefont {Senko}},
  \bibinfo {author} {\bibfnamefont {J.}~\bibnamefont {Smith}}, \bibinfo
  {author} {\bibfnamefont {A.}~\bibnamefont {Lee}}, \bibinfo {author}
  {\bibfnamefont {S.}~\bibnamefont {Korenblit}}, \ and\ \bibinfo {author}
  {\bibfnamefont {C.}~\bibnamefont {Monroe}},\ }\href {\doibase
  10.1103/PhysRevA.88.012334} {\bibfield  {journal} {\bibinfo  {journal} {Phys.
  Rev. A}\ }\textbf {\bibinfo {volume} {88}},\ \bibinfo {pages} {012334}
  (\bibinfo {year} {2013}{\natexlab{b}})}\BibitemShut {NoStop}%
\bibitem [{\citenamefont {Senko}\ \emph {et~al.}(2015)\citenamefont {Senko},
  \citenamefont {Richerme}, \citenamefont {Smith}, \citenamefont {Lee},
  \citenamefont {Cohen}, \citenamefont {Retzker},\ and\ \citenamefont
  {Monroe}}]{Senko2015}%
  \BibitemOpen
  \bibfield  {author} {\bibinfo {author} {\bibfnamefont {C.}~\bibnamefont
  {Senko}}, \bibinfo {author} {\bibfnamefont {P.}~\bibnamefont {Richerme}},
  \bibinfo {author} {\bibfnamefont {J.}~\bibnamefont {Smith}}, \bibinfo
  {author} {\bibfnamefont {A.}~\bibnamefont {Lee}}, \bibinfo {author}
  {\bibfnamefont {I.}~\bibnamefont {Cohen}}, \bibinfo {author} {\bibfnamefont
  {A.}~\bibnamefont {Retzker}}, \ and\ \bibinfo {author} {\bibfnamefont
  {C.}~\bibnamefont {Monroe}},\ }\href {\doibase 10.1103/PhysRevX.5.021026}
  {\bibfield  {journal} {\bibinfo  {journal} {Phys. Rev. X}\ }\textbf {\bibinfo
  {volume} {5}},\ \bibinfo {pages} {021026} (\bibinfo {year}
  {2015})}\BibitemShut {NoStop}%
\bibitem [{\citenamefont {Pedernales}\ \emph {et~al.}(2015)\citenamefont
  {Pedernales}, \citenamefont {Lizuain}, \citenamefont {Felicetti},
  \citenamefont {Romero}, \citenamefont {Lamata},\ and\ \citenamefont
  {Solano}}]{Pedernales2015}%
  \BibitemOpen
  \bibfield  {author} {\bibinfo {author} {\bibfnamefont {J.~S.}\ \bibnamefont
  {Pedernales}}, \bibinfo {author} {\bibfnamefont {I.}~\bibnamefont {Lizuain}},
  \bibinfo {author} {\bibfnamefont {S.}~\bibnamefont {Felicetti}}, \bibinfo
  {author} {\bibfnamefont {G.}~\bibnamefont {Romero}}, \bibinfo {author}
  {\bibfnamefont {L.}~\bibnamefont {Lamata}}, \ and\ \bibinfo {author}
  {\bibfnamefont {E.}~\bibnamefont {Solano}},\ }\href@noop {} {\bibfield
  {journal} {\bibinfo  {journal} {Scientific Reports}\ }\textbf {\bibinfo
  {volume} {5}},\ \bibinfo {pages} {15472} (\bibinfo {year} {2015})},\ \bibinfo
  {note} {article}\BibitemShut {NoStop}%
\bibitem [{\citenamefont {Lv}\ \emph {et~al.}(2017)\citenamefont {Lv},
  \citenamefont {An}, \citenamefont {Liu}, \citenamefont {Zhang}, \citenamefont
  {Pedernales}, \citenamefont {Lamata}, \citenamefont {Solano},\ and\
  \citenamefont {Kim}}]{Lv2017}%
  \BibitemOpen
  \bibfield  {author} {\bibinfo {author} {\bibfnamefont {D.}~\bibnamefont
  {Lv}}, \bibinfo {author} {\bibfnamefont {S.}~\bibnamefont {An}}, \bibinfo
  {author} {\bibfnamefont {Z.}~\bibnamefont {Liu}}, \bibinfo {author}
  {\bibfnamefont {J.-N.}\ \bibnamefont {Zhang}}, \bibinfo {author}
  {\bibfnamefont {J.~S.}\ \bibnamefont {Pedernales}}, \bibinfo {author}
  {\bibfnamefont {L.}~\bibnamefont {Lamata}}, \bibinfo {author} {\bibfnamefont
  {E.}~\bibnamefont {Solano}}, \ and\ \bibinfo {author} {\bibfnamefont
  {K.}~\bibnamefont {Kim}},\ }\href@noop {} {\enquote {\bibinfo {title}
  {Exploring quantum signatures of chaos on a {F}loquet synthetic lattice},}\ }
  (\bibinfo {year} {2017}),\ \Eprint {http://arxiv.org/abs/arXiv:1711.00582}
  {arXiv:1711.00582} \BibitemShut {NoStop}%
\bibitem [{\citenamefont {Johnson}\ \emph {et~al.}(2017)\citenamefont
  {Johnson}, \citenamefont {Wong-Campos}, \citenamefont {Neyenhuis},
  \citenamefont {Mizrahi},\ and\ \citenamefont {Monroe}}]{Johnson2017}%
  \BibitemOpen
  \bibfield  {author} {\bibinfo {author} {\bibfnamefont {K.~G.}\ \bibnamefont
  {Johnson}}, \bibinfo {author} {\bibfnamefont {J.~D.}\ \bibnamefont
  {Wong-Campos}}, \bibinfo {author} {\bibfnamefont {B.}~\bibnamefont
  {Neyenhuis}}, \bibinfo {author} {\bibfnamefont {J.}~\bibnamefont {Mizrahi}},
  \ and\ \bibinfo {author} {\bibfnamefont {C.}~\bibnamefont {Monroe}},\ }\href
  {\doibase 10.1038/s41467-017-00682-6} {\bibfield  {journal} {\bibinfo
  {journal} {Nature Communications}\ }\textbf {\bibinfo {volume} {8}},\
  \bibinfo {pages} {697} (\bibinfo {year} {2017})}\BibitemShut {NoStop}%
\bibitem [{\citenamefont {Kienzler}\ \emph {et~al.}(2016)\citenamefont
  {Kienzler}, \citenamefont {Fl\"uhmann}, \citenamefont {Negnevitsky},
  \citenamefont {Lo}, \citenamefont {Marinelli}, \citenamefont {Nadlinger},\
  and\ \citenamefont {Home}}]{Kienzler2016}%
  \BibitemOpen
  \bibfield  {author} {\bibinfo {author} {\bibfnamefont {D.}~\bibnamefont
  {Kienzler}}, \bibinfo {author} {\bibfnamefont {C.}~\bibnamefont
  {Fl\"uhmann}}, \bibinfo {author} {\bibfnamefont {V.}~\bibnamefont
  {Negnevitsky}}, \bibinfo {author} {\bibfnamefont {H.-Y.}\ \bibnamefont {Lo}},
  \bibinfo {author} {\bibfnamefont {M.}~\bibnamefont {Marinelli}}, \bibinfo
  {author} {\bibfnamefont {D.}~\bibnamefont {Nadlinger}}, \ and\ \bibinfo
  {author} {\bibfnamefont {J.~P.}\ \bibnamefont {Home}},\ }\href {\doibase
  10.1103/PhysRevLett.116.140402} {\bibfield  {journal} {\bibinfo  {journal}
  {Phys. Rev. Lett.}\ }\textbf {\bibinfo {volume} {116}},\ \bibinfo {pages}
  {140402} (\bibinfo {year} {2016})}\BibitemShut {NoStop}%
\bibitem [{\citenamefont {Monroe}\ \emph {et~al.}(1996)\citenamefont {Monroe},
  \citenamefont {Meekhof}, \citenamefont {King},\ and\ \citenamefont
  {Wineland}}]{Monroe1996}%
  \BibitemOpen
  \bibfield  {author} {\bibinfo {author} {\bibfnamefont {C.}~\bibnamefont
  {Monroe}}, \bibinfo {author} {\bibfnamefont {D.~M.}\ \bibnamefont {Meekhof}},
  \bibinfo {author} {\bibfnamefont {B.~E.}\ \bibnamefont {King}}, \ and\
  \bibinfo {author} {\bibfnamefont {D.~J.}\ \bibnamefont {Wineland}},\ }\href
  {\doibase 10.1126/science.272.5265.1131} {\bibfield  {journal} {\bibinfo
  {journal} {Science}\ }\textbf {\bibinfo {volume} {272}},\ \bibinfo {pages}
  {1131} (\bibinfo {year} {1996})}\BibitemShut {NoStop}%
\bibitem [{\citenamefont {Toyoda}\ \emph {et~al.}(2015)\citenamefont {Toyoda},
  \citenamefont {Hiji}, \citenamefont {Noguchi},\ and\ \citenamefont
  {Urabe}}]{Toyoda2015}%
  \BibitemOpen
  \bibfield  {author} {\bibinfo {author} {\bibfnamefont {K.}~\bibnamefont
  {Toyoda}}, \bibinfo {author} {\bibfnamefont {R.}~\bibnamefont {Hiji}},
  \bibinfo {author} {\bibfnamefont {A.}~\bibnamefont {Noguchi}}, \ and\
  \bibinfo {author} {\bibfnamefont {S.}~\bibnamefont {Urabe}},\ }\href@noop {}
  {\bibfield  {journal} {\bibinfo  {journal} {Nature}\ }\textbf {\bibinfo
  {volume} {527}},\ \bibinfo {pages} {74} (\bibinfo {year} {2015})}\BibitemShut
  {NoStop}%
\bibitem [{\citenamefont {Debnath}\ \emph {et~al.}(2017)\citenamefont
  {Debnath}, \citenamefont {Linke}, \citenamefont {Wang}, \citenamefont
  {Figgatt}, \citenamefont {Landsman}, \citenamefont {Duan},\ and\
  \citenamefont {Monroe}}]{Debnath2017}%
  \BibitemOpen
  \bibfield  {author} {\bibinfo {author} {\bibfnamefont {S.}~\bibnamefont
  {Debnath}}, \bibinfo {author} {\bibfnamefont {N.~M.}\ \bibnamefont {Linke}},
  \bibinfo {author} {\bibfnamefont {S.-T.}\ \bibnamefont {Wang}}, \bibinfo
  {author} {\bibfnamefont {C.}~\bibnamefont {Figgatt}}, \bibinfo {author}
  {\bibfnamefont {K.~A.}\ \bibnamefont {Landsman}}, \bibinfo {author}
  {\bibfnamefont {L.-M.}\ \bibnamefont {Duan}}, \ and\ \bibinfo {author}
  {\bibfnamefont {C.}~\bibnamefont {Monroe}},\ }\href@noop {} {\enquote
  {\bibinfo {title} {Observation of hopping and blockade of bosons in a trapped
  ion spin chain},}\ } (\bibinfo {year} {2017}),\ \Eprint
  {http://arxiv.org/abs/arXiv:1711.00216} {arXiv:1711.00216} \BibitemShut
  {NoStop}%
\bibitem [{\citenamefont {Baumann}\ \emph {et~al.}(2010)\citenamefont
  {Baumann}, \citenamefont {Guerlin}, \citenamefont {Brennecke},\ and\
  \citenamefont {Esslinger}}]{Baumann2010}%
  \BibitemOpen
  \bibfield  {author} {\bibinfo {author} {\bibfnamefont {K.}~\bibnamefont
  {Baumann}}, \bibinfo {author} {\bibfnamefont {C.}~\bibnamefont {Guerlin}},
  \bibinfo {author} {\bibfnamefont {F.}~\bibnamefont {Brennecke}}, \ and\
  \bibinfo {author} {\bibfnamefont {T.}~\bibnamefont {Esslinger}},\ }\href
  {\doibase 10.1038/nature09009} {\bibfield  {journal} {\bibinfo  {journal}
  {Nature}\ }\textbf {\bibinfo {volume} {464}},\ \bibinfo {pages} {1301}
  (\bibinfo {year} {2010})}\BibitemShut {NoStop}%
\bibitem [{\citenamefont {Baumann}\ \emph {et~al.}(2011)\citenamefont
  {Baumann}, \citenamefont {Mottl}, \citenamefont {Brennecke},\ and\
  \citenamefont {Esslinger}}]{Baumann2011}%
  \BibitemOpen
  \bibfield  {author} {\bibinfo {author} {\bibfnamefont {K.}~\bibnamefont
  {Baumann}}, \bibinfo {author} {\bibfnamefont {R.}~\bibnamefont {Mottl}},
  \bibinfo {author} {\bibfnamefont {F.}~\bibnamefont {Brennecke}}, \ and\
  \bibinfo {author} {\bibfnamefont {T.}~\bibnamefont {Esslinger}},\ }\href
  {\doibase 10.1103/PhysRevLett.107.140402} {\bibfield  {journal} {\bibinfo
  {journal} {Phys. Rev. Lett.}\ }\textbf {\bibinfo {volume} {107}},\ \bibinfo
  {pages} {140402} (\bibinfo {year} {2011})}\BibitemShut {NoStop}%
\bibitem [{\citenamefont {Klinder}\ \emph {et~al.}(2015)\citenamefont
  {Klinder}, \citenamefont {Keßler}, \citenamefont {Wolke}, \citenamefont
  {Mathey},\ and\ \citenamefont {Hemmerich}}]{Klinder2015}%
  \BibitemOpen
  \bibfield  {author} {\bibinfo {author} {\bibfnamefont {J.}~\bibnamefont
  {Klinder}}, \bibinfo {author} {\bibfnamefont {H.}~\bibnamefont {Keßler}},
  \bibinfo {author} {\bibfnamefont {M.}~\bibnamefont {Wolke}}, \bibinfo
  {author} {\bibfnamefont {L.}~\bibnamefont {Mathey}}, \ and\ \bibinfo {author}
  {\bibfnamefont {A.}~\bibnamefont {Hemmerich}},\ }\href {\doibase
  10.1073/pnas.1417132112} {\bibfield  {journal} {\bibinfo  {journal} {PNAS}\
  }\textbf {\bibinfo {volume} {112}},\ \bibinfo {pages} {3290} (\bibinfo {year}
  {2015})}\BibitemShut {NoStop}%
\bibitem [{\citenamefont {Fink}\ \emph {et~al.}(2009)\citenamefont {Fink},
  \citenamefont {Bianchetti}, \citenamefont {Baur}, \citenamefont {G\"oppl},
  \citenamefont {Steffen}, \citenamefont {Filipp}, \citenamefont {Leek},
  \citenamefont {Blais},\ and\ \citenamefont {Wallraff}}]{Fink2009}%
  \BibitemOpen
  \bibfield  {author} {\bibinfo {author} {\bibfnamefont {J.~M.}\ \bibnamefont
  {Fink}}, \bibinfo {author} {\bibfnamefont {R.}~\bibnamefont {Bianchetti}},
  \bibinfo {author} {\bibfnamefont {M.}~\bibnamefont {Baur}}, \bibinfo {author}
  {\bibfnamefont {M.}~\bibnamefont {G\"oppl}}, \bibinfo {author} {\bibfnamefont
  {L.}~\bibnamefont {Steffen}}, \bibinfo {author} {\bibfnamefont
  {S.}~\bibnamefont {Filipp}}, \bibinfo {author} {\bibfnamefont {P.~J.}\
  \bibnamefont {Leek}}, \bibinfo {author} {\bibfnamefont {A.}~\bibnamefont
  {Blais}}, \ and\ \bibinfo {author} {\bibfnamefont {A.}~\bibnamefont
  {Wallraff}},\ }\href {\doibase 10.1103/PhysRevLett.103.083601} {\bibfield
  {journal} {\bibinfo  {journal} {Phys. Rev. Lett.}\ }\textbf {\bibinfo
  {volume} {103}},\ \bibinfo {pages} {083601} (\bibinfo {year}
  {2009})}\BibitemShut {NoStop}%
\bibitem [{\citenamefont {G\"{a}rttner}\ \emph {et~al.}(2017)\citenamefont
  {G\"{a}rttner}, \citenamefont {Bohnet}, \citenamefont {Safavi-Naini},
  \citenamefont {Wall}, \citenamefont {Bollinger},\ and\ \citenamefont
  {Rey}}]{Martin2017_OTOC}%
  \BibitemOpen
  \bibfield  {author} {\bibinfo {author} {\bibfnamefont {M.}~\bibnamefont
  {G\"{a}rttner}}, \bibinfo {author} {\bibfnamefont {J.~G.}\ \bibnamefont
  {Bohnet}}, \bibinfo {author} {\bibfnamefont {M.}~\bibnamefont
  {Safavi-Naini}}, \bibinfo {author} {\bibfnamefont {M.~L.}\ \bibnamefont
  {Wall}}, \bibinfo {author} {\bibfnamefont {J.~J.}\ \bibnamefont {Bollinger}},
  \ and\ \bibinfo {author} {\bibfnamefont {A.~M.}\ \bibnamefont {Rey}},\ }\href
  {\doibase 10.1038/nphys4119} {\bibfield  {journal} {\bibinfo  {journal} {Nat.
  Phys.}\ }\textbf {\bibinfo {volume} {13}},\ \bibinfo {pages} {781} (\bibinfo
  {year} {2017})}\BibitemShut {NoStop}%
\bibitem [{\citenamefont {Bohnet}\ \emph {et~al.}(2016)\citenamefont {Bohnet},
  \citenamefont {Sawyer}, \citenamefont {Britton}, \citenamefont {Wall},
  \citenamefont {Rey}, \citenamefont {Foss-Feig},\ and\ \citenamefont
  {Bollinger}}]{Bohnet2016}%
  \BibitemOpen
  \bibfield  {author} {\bibinfo {author} {\bibfnamefont {J.~G.}\ \bibnamefont
  {Bohnet}}, \bibinfo {author} {\bibfnamefont {B.~C.}\ \bibnamefont {Sawyer}},
  \bibinfo {author} {\bibfnamefont {J.~W.}\ \bibnamefont {Britton}}, \bibinfo
  {author} {\bibfnamefont {M.~L.}\ \bibnamefont {Wall}}, \bibinfo {author}
  {\bibfnamefont {A.~M.}\ \bibnamefont {Rey}}, \bibinfo {author} {\bibfnamefont
  {M.}~\bibnamefont {Foss-Feig}}, \ and\ \bibinfo {author} {\bibfnamefont
  {J.~J.}\ \bibnamefont {Bollinger}},\ }\href {\doibase
  10.1126/science.aad9958} {\bibfield  {journal} {\bibinfo  {journal}
  {Science}\ }\textbf {\bibinfo {volume} {352}},\ \bibinfo {pages} {1297}
  (\bibinfo {year} {2016})}\BibitemShut {NoStop}%
\bibitem [{\citenamefont {Sawyer}\ \emph
  {et~al.}(2012{\natexlab{a}})\citenamefont {Sawyer}, \citenamefont {Britton},
  \citenamefont {Keith}, \citenamefont {Wang}, \citenamefont {Freericks},
  \citenamefont {Uys}, \citenamefont {Biercuk},\ and\ \citenamefont
  {Bollinger}}]{Sawyer2012}%
  \BibitemOpen
  \bibfield  {author} {\bibinfo {author} {\bibfnamefont {B.~C.}\ \bibnamefont
  {Sawyer}}, \bibinfo {author} {\bibfnamefont {J.~W.}\ \bibnamefont {Britton}},
  \bibinfo {author} {\bibfnamefont {A.~C.}\ \bibnamefont {Keith}}, \bibinfo
  {author} {\bibfnamefont {C.-C.~J.}\ \bibnamefont {Wang}}, \bibinfo {author}
  {\bibfnamefont {J.~K.}\ \bibnamefont {Freericks}}, \bibinfo {author}
  {\bibfnamefont {H.}~\bibnamefont {Uys}}, \bibinfo {author} {\bibfnamefont
  {M.~J.}\ \bibnamefont {Biercuk}}, \ and\ \bibinfo {author} {\bibfnamefont
  {J.~J.}\ \bibnamefont {Bollinger}},\ }\href {\doibase
  10.1103/PhysRevLett.108.213003} {\bibfield  {journal} {\bibinfo  {journal}
  {Phys. Rev. Lett.}\ }\textbf {\bibinfo {volume} {108}},\ \bibinfo {pages}
  {213003} (\bibinfo {year} {2012}{\natexlab{a}})}\BibitemShut {NoStop}%
\bibitem [{\citenamefont {Biercuk}\ \emph {et~al.}(2009)\citenamefont
  {Biercuk}, \citenamefont {Uys}, \citenamefont {Vandevender}, \citenamefont
  {Shiga}, \citenamefont {Itano},\ and\ \citenamefont
  {Bollinger}}]{Biercuk2009}%
  \BibitemOpen
  \bibfield  {author} {\bibinfo {author} {\bibfnamefont {M.~J.}\ \bibnamefont
  {Biercuk}}, \bibinfo {author} {\bibfnamefont {H.}~\bibnamefont {Uys}},
  \bibinfo {author} {\bibfnamefont {A.~P.}\ \bibnamefont {Vandevender}},
  \bibinfo {author} {\bibfnamefont {N.}~\bibnamefont {Shiga}}, \bibinfo
  {author} {\bibfnamefont {W.~M.}\ \bibnamefont {Itano}}, \ and\ \bibinfo
  {author} {\bibfnamefont {J.~J.}\ \bibnamefont {Bollinger}},\ }\href
  {http://dl.acm.org/citation.cfm?id=2012098.2012100} {\bibfield  {journal}
  {\bibinfo  {journal} {Quantum Info. Comput.}\ }\textbf {\bibinfo {volume}
  {9}},\ \bibinfo {pages} {920} (\bibinfo {year} {2009})}\BibitemShut {NoStop}%
\bibitem [{\citenamefont {Wang}\ \emph {et~al.}(2013)\citenamefont {Wang},
  \citenamefont {Keith},\ and\ \citenamefont {Freericks}}]{Wang2013}%
  \BibitemOpen
  \bibfield  {author} {\bibinfo {author} {\bibfnamefont {C.-C.~J.}\
  \bibnamefont {Wang}}, \bibinfo {author} {\bibfnamefont {A.~C.}\ \bibnamefont
  {Keith}}, \ and\ \bibinfo {author} {\bibfnamefont {J.~K.}\ \bibnamefont
  {Freericks}},\ }\href {\doibase 10.1103/PhysRevA.87.013422} {\bibfield
  {journal} {\bibinfo  {journal} {Phys. Rev. A}\ }\textbf {\bibinfo {volume}
  {87}},\ \bibinfo {pages} {013422} (\bibinfo {year} {2013})}\BibitemShut
  {NoStop}%
\bibitem [{\citenamefont {Dicke}(1954)}]{Dicke1954}%
  \BibitemOpen
  \bibfield  {author} {\bibinfo {author} {\bibfnamefont {R.~H.}\ \bibnamefont
  {Dicke}},\ }\href {\doibase 10.1103/PhysRev.93.99} {\bibfield  {journal}
  {\bibinfo  {journal} {Phys. Rev.}\ }\textbf {\bibinfo {volume} {93}},\
  \bibinfo {pages} {99} (\bibinfo {year} {1954})}\BibitemShut {NoStop}%
\bibitem [{\citenamefont {Garraway}(2011)}]{Garraway2011}%
  \BibitemOpen
  \bibfield  {author} {\bibinfo {author} {\bibfnamefont {B.~M.}\ \bibnamefont
  {Garraway}},\ }\href {\doibase 10.1098/rsta.2010.0333} {\bibfield  {journal}
  {\bibinfo  {journal} {Philos. Trans. Royal Soc. A}\ }\textbf {\bibinfo
  {volume} {369}},\ \bibinfo {pages} {1137} (\bibinfo {year}
  {2011})}\BibitemShut {NoStop}%
\bibitem [{\citenamefont {Wall}\ \emph {et~al.}(2017)\citenamefont {Wall},
  \citenamefont {Safavi-Naini},\ and\ \citenamefont {Rey}}]{Wall2017}%
  \BibitemOpen
  \bibfield  {author} {\bibinfo {author} {\bibfnamefont {M.~L.}\ \bibnamefont
  {Wall}}, \bibinfo {author} {\bibfnamefont {A.}~\bibnamefont {Safavi-Naini}},
  \ and\ \bibinfo {author} {\bibfnamefont {A.~M.}\ \bibnamefont {Rey}},\ }\href
  {\doibase 10.1103/PhysRevA.95.013602} {\bibfield  {journal} {\bibinfo
  {journal} {Phys. Rev. A}\ }\textbf {\bibinfo {volume} {95}},\ \bibinfo
  {pages} {013602} (\bibinfo {year} {2017})}\BibitemShut {NoStop}%
\bibitem [{\citenamefont {Emary}\ and\ \citenamefont
  {Brandes}(2003{\natexlab{a}})}]{Emary2003_PRL}%
  \BibitemOpen
  \bibfield  {author} {\bibinfo {author} {\bibfnamefont {C.}~\bibnamefont
  {Emary}}\ and\ \bibinfo {author} {\bibfnamefont {T.}~\bibnamefont
  {Brandes}},\ }\href {\doibase 10.1103/PhysRevLett.90.044101} {\bibfield
  {journal} {\bibinfo  {journal} {Phys. Rev. Lett.}\ }\textbf {\bibinfo
  {volume} {90}},\ \bibinfo {pages} {044101} (\bibinfo {year}
  {2003}{\natexlab{a}})}\BibitemShut {NoStop}%
\bibitem [{\citenamefont {Emary}\ and\ \citenamefont
  {Brandes}(2003{\natexlab{b}})}]{Emary2003_PRE}%
  \BibitemOpen
  \bibfield  {author} {\bibinfo {author} {\bibfnamefont {C.}~\bibnamefont
  {Emary}}\ and\ \bibinfo {author} {\bibfnamefont {T.}~\bibnamefont
  {Brandes}},\ }\href {\doibase 10.1103/PhysRevE.67.066203} {\bibfield
  {journal} {\bibinfo  {journal} {Phys. Rev. E}\ }\textbf {\bibinfo {volume}
  {67}},\ \bibinfo {pages} {066203} (\bibinfo {year}
  {2003}{\natexlab{b}})}\BibitemShut {NoStop}%
\bibitem [{\citenamefont {Ivanov}\ and\ \citenamefont
  {Porras}(2013)}]{Porras2013}%
  \BibitemOpen
  \bibfield  {author} {\bibinfo {author} {\bibfnamefont {P.~A.}\ \bibnamefont
  {Ivanov}}\ and\ \bibinfo {author} {\bibfnamefont {D.}~\bibnamefont
  {Porras}},\ }\href {\doibase 10.1103/PhysRevA.88.023803} {\bibfield
  {journal} {\bibinfo  {journal} {Phys. Rev. A}\ }\textbf {\bibinfo {volume}
  {88}},\ \bibinfo {pages} {023803} (\bibinfo {year} {2013})}\BibitemShut
  {NoStop}%
\bibitem [{\citenamefont {Wunsche}(1991)}]{Wunsche1991}%
  \BibitemOpen
  \bibfield  {author} {\bibinfo {author} {\bibfnamefont {A.}~\bibnamefont
  {Wunsche}},\ }\href@noop {} {\bibfield  {journal} {\bibinfo  {journal}
  {{Quantum Optics: Journal of the European Optical Society Part B}}\ }\textbf
  {\bibinfo {volume} {3}},\ \bibinfo {pages} {359} (\bibinfo {year}
  {1991})}\BibitemShut {NoStop}%
\bibitem [{\citenamefont {Morrison}\ and\ \citenamefont
  {Parkins}(2008)}]{Morrison2008}%
  \BibitemOpen
  \bibfield  {author} {\bibinfo {author} {\bibfnamefont {S.}~\bibnamefont
  {Morrison}}\ and\ \bibinfo {author} {\bibfnamefont {A.~S.}\ \bibnamefont
  {Parkins}},\ }\href {\doibase 10.1103/PhysRevA.77.043810} {\bibfield
  {journal} {\bibinfo  {journal} {Phys. Rev. A}\ }\textbf {\bibinfo {volume}
  {77}},\ \bibinfo {pages} {043810} (\bibinfo {year} {2008})}\BibitemShut
  {NoStop}%
\bibitem [{\citenamefont {Elliott}\ \emph {et~al.}(1970)\citenamefont
  {Elliott}, \citenamefont {Pfeuty},\ and\ \citenamefont {Wood}}]{Elliott1970}%
  \BibitemOpen
  \bibfield  {author} {\bibinfo {author} {\bibfnamefont {R.~J.}\ \bibnamefont
  {Elliott}}, \bibinfo {author} {\bibfnamefont {P.}~\bibnamefont {Pfeuty}}, \
  and\ \bibinfo {author} {\bibfnamefont {C.}~\bibnamefont {Wood}},\ }\href
  {\doibase 10.1103/PhysRevLett.25.443} {\bibfield  {journal} {\bibinfo
  {journal} {Phys. Rev. Lett.}\ }\textbf {\bibinfo {volume} {25}},\ \bibinfo
  {pages} {443} (\bibinfo {year} {1970})}\BibitemShut {NoStop}%
\bibitem [{\citenamefont {Gilmore}\ \emph {et~al.}(2017)\citenamefont
  {Gilmore}, \citenamefont {Bohnet}, \citenamefont {Sawyer}, \citenamefont
  {Britton},\ and\ \citenamefont {Bollinger}}]{Gilmore2017}%
  \BibitemOpen
  \bibfield  {author} {\bibinfo {author} {\bibfnamefont {K.~A.}\ \bibnamefont
  {Gilmore}}, \bibinfo {author} {\bibfnamefont {J.~G.}\ \bibnamefont {Bohnet}},
  \bibinfo {author} {\bibfnamefont {B.~C.}\ \bibnamefont {Sawyer}}, \bibinfo
  {author} {\bibfnamefont {J.~W.}\ \bibnamefont {Britton}}, \ and\ \bibinfo
  {author} {\bibfnamefont {J.~J.}\ \bibnamefont {Bollinger}},\ }\href {\doibase
  10.1103/PhysRevLett.118.263602} {\bibfield  {journal} {\bibinfo  {journal}
  {Phys. Rev. Lett.}\ }\textbf {\bibinfo {volume} {118}},\ \bibinfo {pages}
  {263602} (\bibinfo {year} {2017})}\BibitemShut {NoStop}%
\bibitem [{\citenamefont {Balasubramanian}\ \emph {et~al.}()\citenamefont
  {Balasubramanian}, \citenamefont {Han}, \citenamefont {Yoshimura},\ and\
  \citenamefont {Freericks}}]{Freericks2015}%
  \BibitemOpen
  \bibfield  {author} {\bibinfo {author} {\bibfnamefont {S.}~\bibnamefont
  {Balasubramanian}}, \bibinfo {author} {\bibfnamefont {S.}~\bibnamefont
  {Han}}, \bibinfo {author} {\bibfnamefont {B.~T.}\ \bibnamefont {Yoshimura}},
  \ and\ \bibinfo {author} {\bibfnamefont {J.~K.}\ \bibnamefont {Freericks}},\
  }\href@noop {} {\ }\Eprint {http://arxiv.org/abs/1511.04602}
  {arXiv:1511.04602} \BibitemShut {NoStop}%
\bibitem [{\citenamefont {Sawyer}\ \emph
  {et~al.}(2012{\natexlab{b}})\citenamefont {Sawyer}, \citenamefont {Britton},
  \citenamefont {Keith}, \citenamefont {Wang}, \citenamefont {Freericks},
  \citenamefont {Uys}, \citenamefont {Biercuk},\ and\ \citenamefont
  {Bollinger}}]{Sawyer2012modeAndTempSpec}%
  \BibitemOpen
  \bibfield  {author} {\bibinfo {author} {\bibfnamefont {B.~C.}\ \bibnamefont
  {Sawyer}}, \bibinfo {author} {\bibfnamefont {J.~W.}\ \bibnamefont {Britton}},
  \bibinfo {author} {\bibfnamefont {A.~C.}\ \bibnamefont {Keith}}, \bibinfo
  {author} {\bibfnamefont {C.~C.~J.}\ \bibnamefont {Wang}}, \bibinfo {author}
  {\bibfnamefont {J.~K.}\ \bibnamefont {Freericks}}, \bibinfo {author}
  {\bibfnamefont {H.}~\bibnamefont {Uys}}, \bibinfo {author} {\bibfnamefont
  {M.~J.}\ \bibnamefont {Biercuk}}, \ and\ \bibinfo {author} {\bibfnamefont
  {J.~J.}\ \bibnamefont {Bollinger}},\ }\href@noop {} {\bibfield  {journal}
  {\bibinfo  {journal} {Phys. Rev. Lett.}\ } (\bibinfo {year}
  {2012}{\natexlab{b}})},\ \Eprint {http://arxiv.org/abs/arXiv:1201.4415v1}
  {arXiv:1201.4415v1} \BibitemShut {NoStop}%
\bibitem [{\citenamefont {Uys}\ \emph {et~al.}(2010)\citenamefont {Uys},
  \citenamefont {Biercuk}, \citenamefont {VanDevender}, \citenamefont
  {Ospelkaus}, \citenamefont {Meiser}, \citenamefont {Ozeri},\ and\
  \citenamefont {Bollinger}}]{Uys2010}%
  \BibitemOpen
  \bibfield  {author} {\bibinfo {author} {\bibfnamefont {H.}~\bibnamefont
  {Uys}}, \bibinfo {author} {\bibfnamefont {M.~J.}\ \bibnamefont {Biercuk}},
  \bibinfo {author} {\bibfnamefont {A.~P.}\ \bibnamefont {VanDevender}},
  \bibinfo {author} {\bibfnamefont {C.}~\bibnamefont {Ospelkaus}}, \bibinfo
  {author} {\bibfnamefont {D.}~\bibnamefont {Meiser}}, \bibinfo {author}
  {\bibfnamefont {R.}~\bibnamefont {Ozeri}}, \ and\ \bibinfo {author}
  {\bibfnamefont {J.~J.}\ \bibnamefont {Bollinger}},\ }\href {\doibase
  10.1103/PhysRevLett.105.200401} {\bibfield  {journal} {\bibinfo  {journal}
  {Phys. Rev. Lett.}\ }\textbf {\bibinfo {volume} {105}},\ \bibinfo {pages}
  {200401} (\bibinfo {year} {2010})}\BibitemShut {NoStop}%
\bibitem [{\citenamefont {Cirac}\ and\ \citenamefont
  {Zoller}(1995)}]{Cirac1995}%
  \BibitemOpen
  \bibfield  {author} {\bibinfo {author} {\bibfnamefont {J.~I.}\ \bibnamefont
  {Cirac}}\ and\ \bibinfo {author} {\bibfnamefont {P.}~\bibnamefont {Zoller}},\
  }\href {\doibase 10.1103/PhysRevLett.74.4091} {\bibfield  {journal} {\bibinfo
   {journal} {Phys. Rev. Lett.}\ }\textbf {\bibinfo {volume} {74}},\ \bibinfo
  {pages} {4091} (\bibinfo {year} {1995})}\BibitemShut {NoStop}%
\bibitem [{\citenamefont {Huelga}\ \emph {et~al.}(1997)\citenamefont {Huelga},
  \citenamefont {Macchiavello}, \citenamefont {Pellizzari}, \citenamefont
  {Ekert}, \citenamefont {Plenio},\ and\ \citenamefont {Cirac}}]{Huelga1997}%
  \BibitemOpen
  \bibfield  {author} {\bibinfo {author} {\bibfnamefont {S.~F.}\ \bibnamefont
  {Huelga}}, \bibinfo {author} {\bibfnamefont {C.}~\bibnamefont
  {Macchiavello}}, \bibinfo {author} {\bibfnamefont {T.}~\bibnamefont
  {Pellizzari}}, \bibinfo {author} {\bibfnamefont {A.~K.}\ \bibnamefont
  {Ekert}}, \bibinfo {author} {\bibfnamefont {M.~B.}\ \bibnamefont {Plenio}}, \
  and\ \bibinfo {author} {\bibfnamefont {J.~I.}\ \bibnamefont {Cirac}},\ }\href
  {\doibase 10.1103/PhysRevLett.79.3865} {\bibfield  {journal} {\bibinfo
  {journal} {Phys. Rev. Lett.}\ }\textbf {\bibinfo {volume} {79}},\ \bibinfo
  {pages} {3865} (\bibinfo {year} {1997})}\BibitemShut {NoStop}%
\bibitem [{\citenamefont {Lechner}\ \emph {et~al.}(2016)\citenamefont
  {Lechner}, \citenamefont {Maier}, \citenamefont {Hempel}, \citenamefont
  {Jurcevic}, \citenamefont {Lanyon}, \citenamefont {Monz}, \citenamefont
  {Brownnutt}, \citenamefont {Blatt},\ and\ \citenamefont {Roos}}]{Roos2016}%
  \BibitemOpen
  \bibfield  {author} {\bibinfo {author} {\bibfnamefont {R.}~\bibnamefont
  {Lechner}}, \bibinfo {author} {\bibfnamefont {C.}~\bibnamefont {Maier}},
  \bibinfo {author} {\bibfnamefont {C.}~\bibnamefont {Hempel}}, \bibinfo
  {author} {\bibfnamefont {P.}~\bibnamefont {Jurcevic}}, \bibinfo {author}
  {\bibfnamefont {B.~P.}\ \bibnamefont {Lanyon}}, \bibinfo {author}
  {\bibfnamefont {T.}~\bibnamefont {Monz}}, \bibinfo {author} {\bibfnamefont
  {M.}~\bibnamefont {Brownnutt}}, \bibinfo {author} {\bibfnamefont
  {R.}~\bibnamefont {Blatt}}, \ and\ \bibinfo {author} {\bibfnamefont {C.~F.}\
  \bibnamefont {Roos}},\ }\href {\doibase 10.1103/PhysRevA.93.053401}
  {\bibfield  {journal} {\bibinfo  {journal} {Phys. Rev. A}\ }\textbf {\bibinfo
  {volume} {93}},\ \bibinfo {pages} {053401} (\bibinfo {year}
  {2016})}\BibitemShut {NoStop}%
\bibitem [{\citenamefont {Monroe}\ \emph {et~al.}(2014)\citenamefont {Monroe},
  \citenamefont {Raussendorf}, \citenamefont {Ruthven}, \citenamefont {Brown},
  \citenamefont {Maunz}, \citenamefont {Duan},\ and\ \citenamefont
  {Kim}}]{Monroe2014}%
  \BibitemOpen
  \bibfield  {author} {\bibinfo {author} {\bibfnamefont {C.}~\bibnamefont
  {Monroe}}, \bibinfo {author} {\bibfnamefont {R.}~\bibnamefont {Raussendorf}},
  \bibinfo {author} {\bibfnamefont {A.}~\bibnamefont {Ruthven}}, \bibinfo
  {author} {\bibfnamefont {K.~R.}\ \bibnamefont {Brown}}, \bibinfo {author}
  {\bibfnamefont {P.}~\bibnamefont {Maunz}}, \bibinfo {author} {\bibfnamefont
  {L.-M.}\ \bibnamefont {Duan}}, \ and\ \bibinfo {author} {\bibfnamefont
  {J.}~\bibnamefont {Kim}},\ }\href {\doibase 10.1103/PhysRevA.89.022317}
  {\bibfield  {journal} {\bibinfo  {journal} {Phys. Rev. A}\ }\textbf {\bibinfo
  {volume} {89}},\ \bibinfo {pages} {022317} (\bibinfo {year} {2014})},\
  \Eprint {http://arxiv.org/abs/1208.0391} {arXiv:1208.0391} \BibitemShut
  {NoStop}%
\bibitem [{\citenamefont {Benhelm}\ \emph {et~al.}(2008)\citenamefont
  {Benhelm}, \citenamefont {Kirchmair}, \citenamefont {Roos},\ and\
  \citenamefont {Blatt}}]{Blatt2008}%
  \BibitemOpen
  \bibfield  {author} {\bibinfo {author} {\bibfnamefont {J.}~\bibnamefont
  {Benhelm}}, \bibinfo {author} {\bibfnamefont {G.}~\bibnamefont {Kirchmair}},
  \bibinfo {author} {\bibfnamefont {C.~F.}\ \bibnamefont {Roos}}, \ and\
  \bibinfo {author} {\bibfnamefont {B.}~\bibnamefont {Blatt}},\ }\href@noop {}
  {\bibfield  {journal} {\bibinfo  {journal} {Nature Physics}\ }\textbf
  {\bibinfo {volume} {4}},\ \bibinfo {pages} {463} (\bibinfo {year}
  {2008})}\BibitemShut {NoStop}%
\bibitem [{\citenamefont {Monz}\ \emph {et~al.}(2011)\citenamefont {Monz},
  \citenamefont {Schindler}, \citenamefont {Barreiro}, \citenamefont {Chwalla},
  \citenamefont {Nigg}, \citenamefont {Coish}, \citenamefont {Harlander},
  \citenamefont {H\"ansel}, \citenamefont {Hennrich},\ and\ \citenamefont
  {Blatt}}]{PhysRevLett.106.130506}%
  \BibitemOpen
  \bibfield  {author} {\bibinfo {author} {\bibfnamefont {T.}~\bibnamefont
  {Monz}}, \bibinfo {author} {\bibfnamefont {P.}~\bibnamefont {Schindler}},
  \bibinfo {author} {\bibfnamefont {J.~T.}\ \bibnamefont {Barreiro}}, \bibinfo
  {author} {\bibfnamefont {M.}~\bibnamefont {Chwalla}}, \bibinfo {author}
  {\bibfnamefont {D.}~\bibnamefont {Nigg}}, \bibinfo {author} {\bibfnamefont
  {W.~A.}\ \bibnamefont {Coish}}, \bibinfo {author} {\bibfnamefont
  {M.}~\bibnamefont {Harlander}}, \bibinfo {author} {\bibfnamefont
  {W.}~\bibnamefont {H\"ansel}}, \bibinfo {author} {\bibfnamefont
  {M.}~\bibnamefont {Hennrich}}, \ and\ \bibinfo {author} {\bibfnamefont
  {R.}~\bibnamefont {Blatt}},\ }\href {\doibase 10.1103/PhysRevLett.106.130506}
  {\bibfield  {journal} {\bibinfo  {journal} {Phys. Rev. Lett.}\ }\textbf
  {\bibinfo {volume} {106}},\ \bibinfo {pages} {130506} (\bibinfo {year}
  {2011})}\BibitemShut {NoStop}%
\bibitem [{\citenamefont {Braunstein}\ and\ \citenamefont
  {Caves}(1994)}]{Braunstein1994}%
  \BibitemOpen
  \bibfield  {author} {\bibinfo {author} {\bibfnamefont {S.~L.}\ \bibnamefont
  {Braunstein}}\ and\ \bibinfo {author} {\bibfnamefont {C.~M.}\ \bibnamefont
  {Caves}},\ }\href {\doibase 10.1103/PhysRevLett.72.3439} {\bibfield
  {journal} {\bibinfo  {journal} {Phys. Rev. Lett.}\ }\textbf {\bibinfo
  {volume} {72}},\ \bibinfo {pages} {3439} (\bibinfo {year}
  {1994})}\BibitemShut {NoStop}%
\bibitem [{\citenamefont {Boixo}\ \emph {et~al.}(2007)\citenamefont {Boixo},
  \citenamefont {Flammia}, \citenamefont {Caves},\ and\ \citenamefont
  {Geremia}}]{Boixo_2007}%
  \BibitemOpen
  \bibfield  {author} {\bibinfo {author} {\bibfnamefont {S.}~\bibnamefont
  {Boixo}}, \bibinfo {author} {\bibfnamefont {S.~T.}\ \bibnamefont {Flammia}},
  \bibinfo {author} {\bibfnamefont {C.~M.}\ \bibnamefont {Caves}}, \ and\
  \bibinfo {author} {\bibfnamefont {J.}~\bibnamefont {Geremia}},\ }\href
  {\doibase 10.1103/PhysRevLett.98.090401} {\bibfield  {journal} {\bibinfo
  {journal} {Phys. Rev. Lett.}\ }\textbf {\bibinfo {volume} {98}},\ \bibinfo
  {pages} {090401} (\bibinfo {year} {2007})}\BibitemShut {NoStop}%
\bibitem [{\citenamefont {Macr\`{\i}}\ \emph {et~al.}(2016)\citenamefont
  {Macr\`{\i}}, \citenamefont {Smerzi},\ and\ \citenamefont
  {Pezz\`e}}]{Macri_2016}%
  \BibitemOpen
  \bibfield  {author} {\bibinfo {author} {\bibfnamefont {T.}~\bibnamefont
  {Macr\`{\i}}}, \bibinfo {author} {\bibfnamefont {A.}~\bibnamefont {Smerzi}},
  \ and\ \bibinfo {author} {\bibfnamefont {L.}~\bibnamefont {Pezz\`e}},\ }\href
  {\doibase 10.1103/PhysRevA.94.010102} {\bibfield  {journal} {\bibinfo
  {journal} {Phys. Rev. A}\ }\textbf {\bibinfo {volume} {94}},\ \bibinfo
  {pages} {010102} (\bibinfo {year} {2016})}\BibitemShut {NoStop}%
\bibitem [{\citenamefont {Davis}\ \emph {et~al.}(2016)\citenamefont {Davis},
  \citenamefont {Bentsen},\ and\ \citenamefont {Schleier-Smith}}]{Davis_2016}%
  \BibitemOpen
  \bibfield  {author} {\bibinfo {author} {\bibfnamefont {E.}~\bibnamefont
  {Davis}}, \bibinfo {author} {\bibfnamefont {G.}~\bibnamefont {Bentsen}}, \
  and\ \bibinfo {author} {\bibfnamefont {M.}~\bibnamefont {Schleier-Smith}},\
  }\href {\doibase 10.1103/PhysRevLett.116.053601} {\bibfield  {journal}
  {\bibinfo  {journal} {Phys. Rev. Lett.}\ }\textbf {\bibinfo {volume} {116}},\
  \bibinfo {pages} {053601} (\bibinfo {year} {2016})}\BibitemShut {NoStop}%
\bibitem [{\citenamefont {Linnemann}\ \emph {et~al.}(2016)\citenamefont
  {Linnemann}, \citenamefont {Strobel}, \citenamefont {Muessel}, \citenamefont
  {Schulz}, \citenamefont {Lewis-Swan}, \citenamefont {Kheruntsyan},\ and\
  \citenamefont {Oberthaler}}]{Linnemann_2016}%
  \BibitemOpen
  \bibfield  {author} {\bibinfo {author} {\bibfnamefont {D.}~\bibnamefont
  {Linnemann}}, \bibinfo {author} {\bibfnamefont {H.}~\bibnamefont {Strobel}},
  \bibinfo {author} {\bibfnamefont {W.}~\bibnamefont {Muessel}}, \bibinfo
  {author} {\bibfnamefont {J.}~\bibnamefont {Schulz}}, \bibinfo {author}
  {\bibfnamefont {R.~J.}\ \bibnamefont {Lewis-Swan}}, \bibinfo {author}
  {\bibfnamefont {K.~V.}\ \bibnamefont {Kheruntsyan}}, \ and\ \bibinfo {author}
  {\bibfnamefont {M.~K.}\ \bibnamefont {Oberthaler}},\ }\href {\doibase
  10.1103/PhysRevLett.117.013001} {\bibfield  {journal} {\bibinfo  {journal}
  {Phys. Rev. Lett.}\ }\textbf {\bibinfo {volume} {117}},\ \bibinfo {pages}
  {013001} (\bibinfo {year} {2016})}\BibitemShut {NoStop}%
\bibitem [{\citenamefont {Hosten}\ \emph {et~al.}(2016)\citenamefont {Hosten},
  \citenamefont {Krishnakumar}, \citenamefont {Engelsen},\ and\ \citenamefont
  {Kasevich}}]{Hosten_2016}%
  \BibitemOpen
  \bibfield  {author} {\bibinfo {author} {\bibfnamefont {O.}~\bibnamefont
  {Hosten}}, \bibinfo {author} {\bibfnamefont {R.}~\bibnamefont
  {Krishnakumar}}, \bibinfo {author} {\bibfnamefont {N.~J.}\ \bibnamefont
  {Engelsen}}, \ and\ \bibinfo {author} {\bibfnamefont {M.~A.}\ \bibnamefont
  {Kasevich}},\ }\href {\doibase 10.1126/science.aaf3397} {\bibfield  {journal}
  {\bibinfo  {journal} {Science}\ }\textbf {\bibinfo {volume} {352}},\ \bibinfo
  {pages} {1552} (\bibinfo {year} {2016})}\BibitemShut {NoStop}%
\bibitem [{\citenamefont {Meier}\ \emph {et~al.}(2017)\citenamefont {Meier},
  \citenamefont {Ang'ong'a}, \citenamefont {An},\ and\ \citenamefont
  {Gadway}}]{Meier_2017}%
  \BibitemOpen
  \bibfield  {author} {\bibinfo {author} {\bibfnamefont {E.~J.}\ \bibnamefont
  {Meier}}, \bibinfo {author} {\bibfnamefont {J.}~\bibnamefont {Ang'ong'a}},
  \bibinfo {author} {\bibfnamefont {F.~A.}\ \bibnamefont {An}}, \ and\ \bibinfo
  {author} {\bibfnamefont {B.}~\bibnamefont {Gadway}},\ }\href@noop {}
  {\enquote {\bibinfo {title} {Exploring quantum signatures of chaos on a
  floquet synthetic lattice},}\ } (\bibinfo {year} {2017}),\ \Eprint
  {http://arxiv.org/abs/arXiv:1705.06714} {arXiv:1705.06714} \BibitemShut
  {NoStop}%
\end{thebibliography}%

\appendix

\section{Additional sequence to prepare the spin cat-state \label{app:cat}}
In the main text, we briefly outline a procedure to prepare a pure spin-cat state via preparation of the ground-state of the Dicke Hamiltonian. Here, we expand upon this discussion and give the appropriate details to verify this step. 

In the weak-field limit, $B \ll g_0^2 /|\delta|$, the ground-state of the Dicke Hamiltonian is the spin-phonon cat-state: 
\begin{equation}
\vert \psi_{{\rm SB},\, I}^{GS} \rangle = \frac{1}{\sqrt{2}}\Big(\vert\alpha_0,0\rangle\vert N/2 \rangle_z + \vert-\alpha_0,0\rangle\vert -N/2 \rangle_z\Big) , \label{eqn:Supp_SpinPhCat}
\end{equation}
where $\alpha_0 = g_0\sqrt{N}/(2\delta)$. Here $\vert\alpha,n\rangle \equiv \hat{D}(\alpha)\vert n\rangle$ is a displaced Fock state, obtained by acting the displacement operator $\hat{D}(\alpha) = e^{\alpha\hat{a}^{\dagger} - \alpha^*\hat{a}}$ on the Fock state $|n\rangle$. 

Since the spin and phonon degrees of freedom are entangled in the ground-state [Eq.~(\ref{eqn:Supp_SpinPhCat})], the state obtained by tracing over the phonon degree of freedom is characterized by the density operator
\begin{multline}
 \hat{\rho}_s = \frac{1}{2}\Big[ |N/2\rangle_z\langle N/2|_z + |-N/2\rangle_z\langle -N/2|_z \Big] \notag \\
  + \frac{e^{-|\alpha_0|^2}}{2} \Big[|-N/2\rangle_z\langle N/2|_z + |N/2\rangle_z\langle -N/2|_z \Big] .
\end{multline}
As the displacement amplitude $|\alpha_0|$ is increased, the reduced density matrix exponentially loses any information about the coherences which are exhibited in the spin-phonon superposition state. As a concrete example, the ground-states of the main text typically have a mean phonon occupation $|\alpha_0|^2 \sim 2$--$30$, leading to $e^{-|\alpha_0|^2}\lesssim 0.1$. To fully probe the available coherences via only the spin degree of freedom, we must first transform Eq.~(\ref{eqn:Supp_SpinPhCat}) to a spin and phonon product state, 
\begin{equation}
\vert \psi_{{\rm SB}} \rangle = \vert \phi\rangle \otimes \frac{1}{\sqrt{2}} \Big(\vert N/2 \rangle_z + \vert -N/2 \rangle_z\Big) , \label{eqn:Supp_SpinPhProduct}
\end{equation}
where $|\phi\rangle$ is some arbitrary state characterizing the phonon degree of freedom. 

A possible procedure to achieve this decomposition is the following: At the conclusion of the ramp protocol, we fix the transverse field at $B=0$ and quench the detuning $\delta \rightarrow \delta^{\prime} = 2\delta$. The 
spin-phonon state is then allowed to evolve for a duration $t_{d} = \pi/\delta^{\prime}$.
In the interaction picture, the initial spin-phonon superposition state evolves as
\begin{eqnarray}
 \vert \psi_{{\rm SB}} \rangle =  \hat{U}(t) \vert \psi_{{\rm SB},\, I}^{GS} \rangle ,
\end{eqnarray}
where 
\begin{eqnarray}
  \hat{U}(t) & = & \hat{U}_{SB}(t)\hat{U}_{SS}(t) , \\
  \hat{U}_{\rm SS}(t) & = & \exp\left( -i \frac{J}{N} \hat S_z^2 t\right) , \\
 \hat{U}_{\rm SB}(t)& = & \hat D( \beta(t,\delta') S_z) .
 \end{eqnarray}
Here, $\hat U(t)$ is the propagator corresponding to the Dicke Hamiltonian with $B=0$ [Eq.~\eqref{eq:HI} of the main text]. The propagator is comprised of two parts, the spin-spin propagator $\hat U_{\rm SS}(t)$ and the spin-phonon propagator $\hat U_{\rm SP}(t)$ where $\beta(t,\delta)= -g_0(1-e^{-i \delta t})/(2\delta\sqrt{N})$ (see \cite{Wall2017} for a more detailed discussion). 

If at the end of the ramp we quench the detuning to $\delta^{\prime} = 2\delta$ and apply $\hat U(t)$ for $t_d=\pi/\delta^{\prime}$, such that $\beta(t_d,\delta')=-g N/(2\delta)$, it is then clear that $\hat{U}_{SB}$ will displace the phonon coherent states (in a direction dependent on the sign of the $S_z$ component) back to vacuum, $|\pm\alpha_0,0\rangle \rightarrow |0\rangle$. Note that the action of $\hat{U}_{SS}$ on the spin component of the ground-state imprints an irrelevant global phase $\varphi =  JNt_d/2$ on the decoupled state Eq.~(\ref{eqn:Supp_SpinPhProduct}).


An alternative, but closely related, procedure to disentangle the spin-phonon state is to drive the spin-phonon coupling on resonance, $\delta \to \delta^{\prime} = 0$. In this case, 
one must shift the phase of the drive by $\pi/2$ such that the spin-phonon coupling transforms as $\frac{g_0}{\sqrt{N}}(\hat{a} + \hat{a}^{\dagger})\hat{S}_z \to \frac{ig_0}{\sqrt{N}}(\hat{a} - \hat{a}^{\dagger})\hat{S}_z$, and subsequently evolve the system for a duration 
duration $t_d = 1/|\delta|$. Following this procedure results in a spin-dependent coherent displacement of the phonon state back to vacuum, $|\pm\alpha_0,0\rangle \rightarrow |0\rangle$, in a manner similar to the previously discussed protocol. 

We make one further point regarding the disentangling protocols. In the experimental system we generally characterize the initial state of the phonons as a thermal ensemble $\hat{\rho}_{\bar n}$ while the spin-degree of freedom is prepared in a pure state, such that the initial spin-phonon state is $\hat{\rho}_{SB}(0) = \hat{\rho}_{\bar n} \otimes \vert -N/2 \rangle_x \langle -N/2 \vert_x $. If the protocol is adiabatic and there is no coupling between the excited energy levels, then not only is the ground-state component of this initial ensemble mapped to the weak-field ground-state of the Dicke Hamiltonian, but the excited fraction due to the thermal distribution is also mapped identically. This implies that the final state at the end of the ramp protocol will be a mixture of the true ground-state and the low-lying excitations, which, if $\delta^2 < g^2 N$, can be characterised as displaced Fock states $\vert \pm\alpha_0, n \rangle$ where $n$ corresponds to the number of phonon excitations above the true ground-state [these excitations are illustrated in Figs.~1~(a) and (b) of the main text]. 

The action of the hold protocol on these states is to identically displace the phonon state such that $\vert \pm\alpha_0, n \rangle \rightarrow |n\rangle$. This maps the spin-phonon excited states to the form of a product state identical to Eq.~(\ref{eqn:Supp_SpinPhProduct}). Hence, tracing the phonons out of these excited states also recovers the spin cat-state. It is this realization which motivates the relation $\mathcal{F}^{S}_{\mathrm{cat}} \geq \mathcal{F}^{SB}_{GS}$ as a lower bound in the previous section, as it possible for excited spin-phonon states to contribute to an increased fidelity to the pure spin cat-state after the disentangling protocol.

\section{Experimental Optimisation of ramp protocols \label{app:RampOpt}}
To experimentally optimize the ramp protocols demonstrated in this work, we chose to optimize with respect to the total magnetization $\langle |\hat{S}_z| \rangle$ at the end of the ramp. For the EXP ramp, we compared approximately $20$ different ramp profiles that utilized different exponential decay rates. Specifically, we would perform an experiment where the effective transverse field was ramped from the initial field $B(t=0)$ at a fixed decay rate to $B\approx0$, where we then measured the spin-projection $M_z^{\mathrm{exp}}$ along the $\hat{z}$-axis. This experiment was repeated, typically $500-700$ times, to gather statistics on the resulting distribution and obtain a measurement of $\langle|\hat{S}_z|\rangle$ from the histogram of $M_z^{\mathrm{exp}}$ measurements.  We then picked a ramp profile with a different exponential decay rate, and repeated this procedure. After identifying the exponential decay rate that optimized the final magnetization $\langle|\hat{S}_z|\rangle$, we performed experiments that measured the magnetization distribution $P(M_z^{\mathrm{exp}})$ when stopping the ramp at different times, as discussed in the main text. We followed a similar procedure to experimentally optimize the LAA ramp profile, where approximately $10$ different ramp profiles were calculated using different spin-spin interaction strength $J$.  We consistently found that the ramp profile calculated from the independently measured spin-spin interaction $J/(2\pi)=1.75$~kHz achieved the maximum observed $\langle|\hat{S}_z|\rangle$.

\begin{figure}[!]
    \includegraphics[width=8cm]{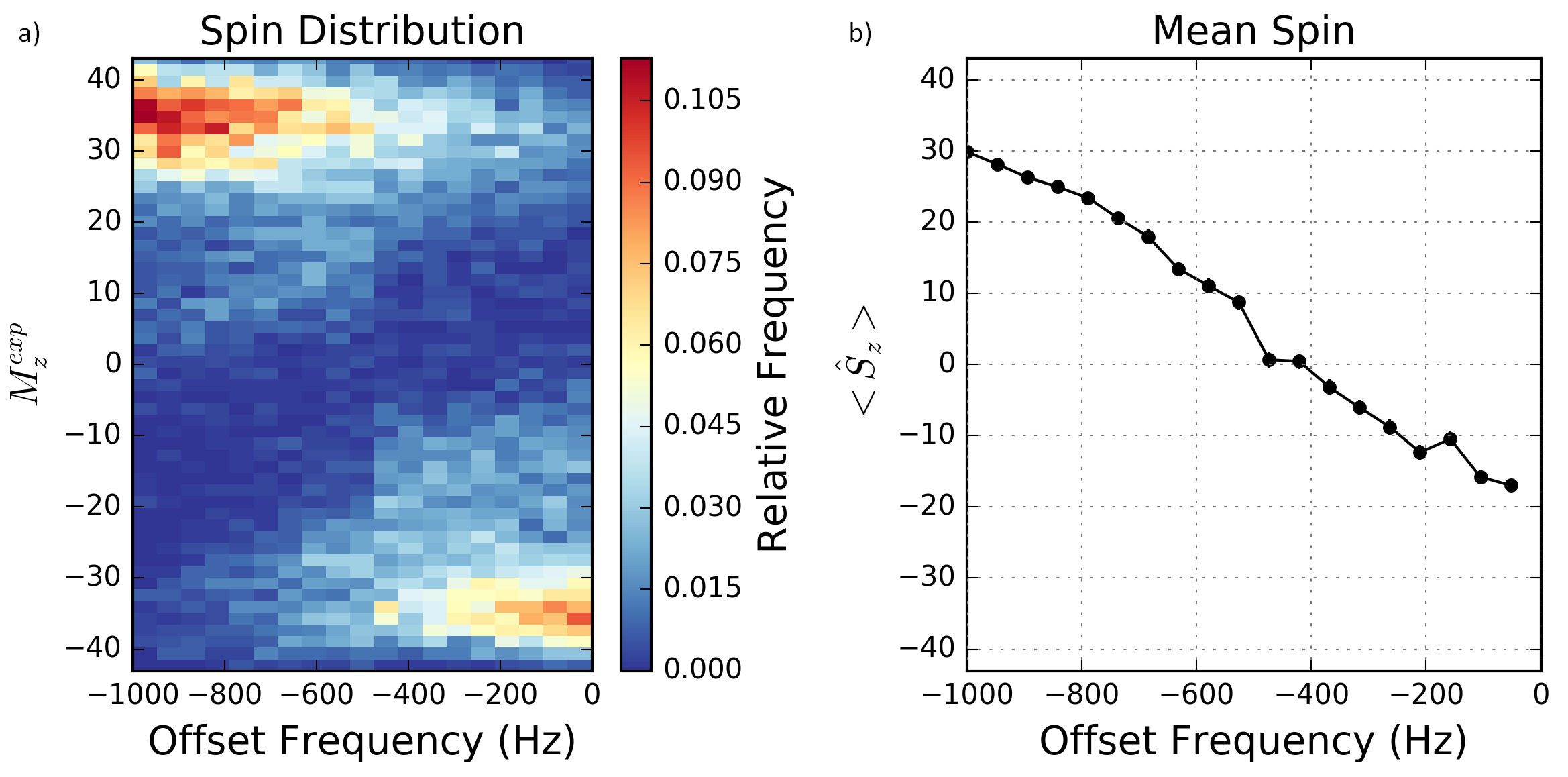}
    \caption{Balancing the $P(M_z^{\mathrm{exp}})$ distributions. (a) $P(M_z^{\mathrm{exp}})$ distribution functions extracted from experimental measurements of the spin-projection $M_z^{\mathrm{exp}}$ at the end of an EXP ramp of the transverse magnetic field to zero. The distribution functions are plotted as a function of frequency offset of the microwaves that generate the effective transverse magnetic field from the spin-flip resonance in the absence of the spin-dependent force. (b) Plot of the average magnetization $\langle\hat{S}_z\rangle$ from (a) as a function of the microwave offset frequency. An offset frequency that balanced the distributions at the end of the ramping sequence, defined by $\langle\hat{S}_z\rangle$, was used in studies described in the main text that measured the spin-projection distribution when stopping the ramp at different times.}
    \label{fig:Supp_Exp}
\end{figure}

When performing these ramp sequences and observing the distributions of $M_z^{\mathrm{exp}}$, in some cases the distributions would be biased to positive or negative spin-projection. This can be observed in the distribution of Fig.~\ref{fig:Supp_Exp}(a) at zero offset frequency. Such an effect can be explained by a small longitudinal magnetic field that breaks the symmetry of the ground state. We discuss the resilience of our typical protocol to such stray fields in the next section. The small longitudinal field was likely due to imperfect nulling of the Stark shift from the off-resonant laser beams that generate the spin-dependent force \cite{Bohnet2016}. We would observe that this effect varies day to day. To compensate for this effect, during the ramp we would apply a small frequency offset to the microwaves that provided the effective transverse field. For each frequency offset, we would measure the distribution of measurements $M_z^{\mathrm{exp}}$ at the end of the transverse field ramp as shown in Fig.~\ref{fig:Supp_Exp}(a). For the appropriate offset, the distribution would be balanced, with large, separated peaks at positive and negative values of $M_z^{\mathrm{exp}}$.  To choose the optimum, we plot $\langle\hat{S}_z\rangle$ as a function of the frequency offset and extract the zero crossing, as shown in Fig.~\ref{fig:Supp_Exp}(b).

\section{Effects of residual longitudinal fields}

In the presence of a non-zero longitudinal field $B_z$ the Hamiltonian describing the system is given by, 
\begin{eqnarray}
  \hat{H}(t) & = & -\delta \hat{a}^{\dagger} \hat{a} + \frac{g_0}{\sqrt{N}}\left(\hat{a} + \hat{a}^{\dagger}\right) \hat{S}_z \notag \\
  & & + B(t)\hat{S}_x + B_z\hat{S}_z .
\end{eqnarray}
The ground-state of this Hamiltonian is no longer degenerate, and is given by one of the two states, $\vert \pm\alpha_0\rangle \otimes \vert \pm N/2 \rangle_z$, depending on the sign of the additional longitudinal field. 

Furthermore, this Hamiltonian is no longer symmetric under the transformation $\hat{S}_z \to -\hat{S}_z$, $\hat{S}_y \to -\hat{S}_y$ and $\hat{a} \to -\hat{a}$. In principle, this loss of symmetry implies that the ground-state at large $B(t)$ given by $\vert \psi^{GS}_{SB, B}\rangle$ is no longer restricted to adiabatically connect to the cat ground-state $\vert \psi^{GS}_{SB, I}\rangle$. Instead it preferentially connects to one of the two states $\vert \pm\alpha_0\rangle \otimes \vert \pm N/2 \rangle_z$, depending on which has the lower energy for a given $B_z$. 
 
\begin{figure}
    \includegraphics[width=0.45\textwidth]{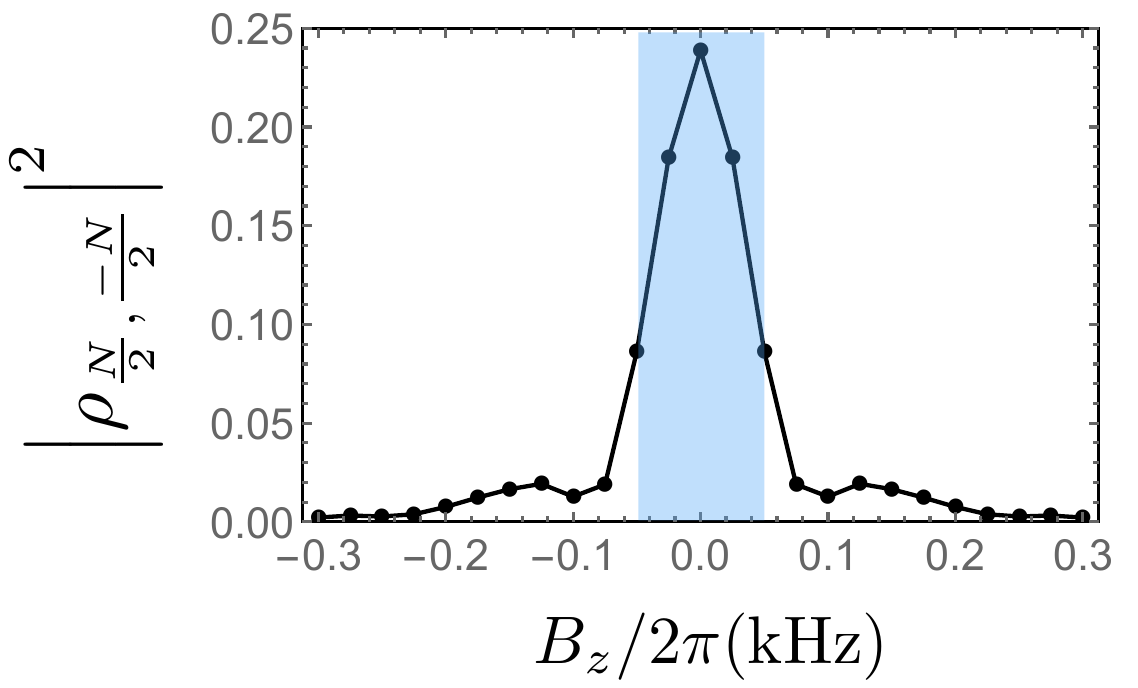}
    \caption{The effect of residual longitudinal fields on off-diagonal coherences. For residual fields of $B_z/(2\pi) \lesssim 50$~Hz, indicated by the shaded region, the state at the end of the EXP protocol contains appreciable coherence between the $\vert \frac N2\rangle_z$ and $\vert -\frac N2\rangle_z$ states.    }
    \label{fig:bzscan}
\end{figure}
 
Despite this, we find that in the presence of a moderate symmetry breaking field $B_z \neq 0$ we can still prepare a state which has significant overlap with the cat-state. We characterize the resilience of our protocol to finite $B_z$ by computing the magnitude of the extremal off-diagonal coherences of the reduced density matrix of the spins following the spin-phonon disentangling protocol discussed in Appendix~\ref{app:cat}. We consider an EXP ramp with $\tau_{\rm ramp}=2$~ms. Following the disentangling procedure, and after tracing out the phonons, the density matrix of the spins can be decomposed as $\hat \rho = \sum_{m,n} \rho_{m,n} \vert m \rangle \langle n\vert$. The coherences, plotted in Fig.~\ref{fig:bzscan}, are then given by the coefficients $\rho_{\frac{N}{2},-\frac{N}{2}}$ and $\rho_{-\frac{N}{2},\frac{N}{2}}$. We find the protocol still achieves significant coherence for $B_z/(2\pi)\lesssim 50$~Hz, which is a level of control over stray longitudinal fields which is experimentally achievable.

We attribute the maintained coherence to the non-adiabaticity of the ramp with respect to the gap, in the limit  $B_x \to 0$ and a small $B_z$ field, between the states $\vert \frac N 2 \rangle_z$ and $\vert -\frac N 2 \rangle_z$ given by $\Delta = B_z N$. Specifically, if the ramp is short such that $\tau_{\mathrm{ramp}} \lesssim 1/(B_z N)$ then diabatic excitations couple the states and we retain a coherent superposition. Extending this more generally, to achieve significant cat-state fidelity we require $1/J \lesssim \tau_{\mathrm{ramp}} \lesssim 1/(B_z N)$ such that the ramp generates diabatic excitations at $B_x \to 0$, yet remains adiabatic with respect to the energy gap at the critical point $B_x \sim J$.

\end{document}